\documentclass[fleqn,usenatbib]{mnras}
\usepackage{graphicx}	
\usepackage{amsmath}	
\usepackage{amssymb}
\usepackage{float}
\usepackage{fix-cm}
\usepackage{subfig}
\usepackage{enumitem}
\usepackage{longtable}
\usepackage{hyperref}
\usepackage{placeins}
\usepackage{subcaption}
\usepackage{stfloats}
\usepackage[most]{tcolorbox}

\title[Identifying warped galaxies using DCNN]{Identifying Warped Galaxies in Pan-STARRS and Euclid using Deep Convolutional Neural Network}

\author[Saranya et al.]{
Saranya Suguna,\thanks{E-mail: saranya.suguna5@gmail.com}
 and Arunima Banerjee
\\
Indian Institute of Science Education and Research, Tirupati 517619, India\\
}

\date{Accepted XXX. Received YYY; in original form ZZZ}

\pubyear{\the\year{}}

\begin{document}
\label{firstpage}
\pagerange{\pageref{firstpage}--\pageref{lastpage}}
\maketitle

\begin{abstract}

Warped galactic discs are common, yet their detection remains challenging, as the outskirts of galaxies are typically faint. Advances in deep imaging surveys improve the detectability of such features, while machine learning enables efficient analysis of large datasets. Using the Pan-STARRS EGIPS catalogue, we develop a deep learning framework by finetuning the Zoobot convNext-nano model on 1000 edge-on galaxy FITS images to distinguish warped and non-warped edge-on galaxies with 83\% accuracy. The trained model is then applied to a larger sample, identifying 2088 warped and 1398 non-warped galaxies with a high prediction probability threshold ($\geq 0.85$). Additionally, we use the model to predict on 3226 edge-on galaxies from the Euclid Q1 survey, demonstrating the model's ability to generalise across datasets with differing resolutions. To analyse the model predictions, we employ LayerCAM to identify the regions of galaxy images that contribute to the classification. We find that warped galaxies differ primarily in their structural properties, exhibiting lower axis ratios and higher asymmetry. Warped galaxies were found to be bluer, with younger stellar populations and enhanced star formation. These results highlight the effectiveness of deep learning methods in identifying subtle morphological features, such as warps, and demonstrate their potential for studying structural properties of galaxies in current and upcoming large imaging surveys.

\end{abstract}

\begin{keywords}
galaxies: structure – galaxies: evolution – cosmology: observations – methods: data analysis
\end{keywords}

%%%%%%%%%%%%%%%%%%%%%%%%%%%%%%%%%%%%%%%%%%%%%%%%%%

%%%%%%%%%%%%%%%%% BODY OF PAPER %%%%%%%%%%%%%%%%%%

\section{Introduction}

Warped discs are morphological features arising from the vertical deviation of the outer regions of the galaxy relative to the inner disc. Observations indicate that nearly 50\% of nearby disc galaxies exhibit appreciable warps in both their stellar and atomic hydrogen (H\,\textsc{i}) discs \citep{1998A&A...337....9R, 2003A&A...399..457S, 2016MNRAS.461.4233R}. The warped disc of the Milky Way has been studied using various tracers \citep{1957AJ.....62...90B, 1957Natur.180..677K, 1965MNRAS.129..299L, 1990ApJ...352...15B, 2006ApJ...643..881L, 2017A&A...602A..67A, 2022A&A...664A..58C, 2023A&A...673A..99U}. In disk dynamical theories, warps are described as vertical bending modes, dominated by m=0 (U-shaped) and m=1 (S-shaped) components in a Fourier decomposition of the vertical displacement of the disc, along with asymmetric warps also commonly observed \citep{2006A&A...446..897S, 2014ApJ...789...90K, 2017MNRAS.472.2751C, 2022ApJ...935...48Z, 2023MNRAS.526.1138G}. The vertical structure of galactic discs has also been studied using IllustrisTNG \citep{2020MNRAS.498.3535S, 2023MNRAS.523.3915S} and other numerical simulations \citep{2020MNRAS.497.2039M, 2024MNRAS.533.2997W, 2023MNRAS.518.5403G}. Warps trace the dynamical coupling between the disc, dark matter halo, satellites, and the surrounding environment, thereby offering key insights into the processes driving galaxy evolution. See \citet{1992ARA&A..30...51B, 1990ApJ...352...15B, 1988gera.book..295B, 2013pss5.book..923S} for general reviews on warps. 

Several theories have been proposed to explain the origin of warps in disc galaxies, which include warp excitation by tidal interactions with companion galaxies such as flyby \citep{2014ApJ...789...90K}, accretion \citep{1997ApJ...480..503H, 2003Ap&SS.284..747B}, mergers \citep{2001A&A...373..402S, 2021MNRAS.508..541P}. A misaligned dark matter halo can perturb the disc and induce warping \citep{1995MNRAS.275..897N, 1999ApJ...513L.107D, 2023ApJ...957L..24H}. Internally driven warps have been investigated by \citet{1988MNRAS.234..873S, 2022MNRAS.510.1375S}. The shape of the dark matter halo can also influence the formation of warps \citep{2000MNRAS.311..733I, 2009ApJ...696.1899J}. Additional contributions from cosmic infall \citep{2006MNRAS.370....2S, 2002A&A...386..169L} and intergalactic magnetic fields \citep{1998A&A...332..809B, 2006MNRAS.365..555S} have also been studied. However, the dominant mechanism responsible for warp formation remains debated. Further, warped galaxies have been studied using a range of large imaging surveys. \citet{2016JKAS...49..239A} investigated the environmental dependence of warps in spiral galaxies from Sloan Digital Sky Survey (SDSS), similarly \citet{2022ApJ...935...48Z} studied U-shaped warps in group/cluster environments. Infrared observations from the Spitzer Space Telescope have been used by \citet{2009MNRAS.396..409S} to study the dynamical origin of warps and their connection to disc–halo interactions. \citet{2025arXiv251109518W} used DESI Legacy Imaging Surveys to study warps. High-resolution imaging from the Hubble Space Telescope (HST) and the James Webb Space Telescope (JWST) has enabled the study of warped galaxies at higher redshifts, including analyses of their redshift distributions and evolution \citep{2025A&A...697L...1R}.

Machine learning techniques have become increasingly effective for analysing galaxy structure and morphology in large imaging datasets. Early applications of Artificial Neural Networks (ANNs) include star–galaxy separation and reproducing galaxy morphologies \citep{2010MNRAS.406..342B}. With the advent of deep learning, Deep Convolutional Neural Networks (DCNNs) have significantly improved the identification of complex morphological features in galaxy images \citep{2015MNRAS.450.1441D}. These methods have been applied to a various of galaxy morphologies, including the detection of barred galaxies \citep{2018MNRAS.477..894A}, interacting systems \citep{2020MNRAS.497.3323P}, spiral structure classification using architectures such as AlexNet \citep{2023MNRAS.518.1022S}, object detection using YOLOv5 \citep{2025PASP..137c4101C}, galaxy merger identification \citep{2026ApJ...998..331L}, and segmentation approaches such as U-Net \citep{2025A&A...702A.258D}. \citet{2025arXiv251109518W} applied deep learning to identify warped galaxies in the DESI Legacy Survey. Transfer learning is a method in which models pretrained on large, general image datasets are adopted for more specialised tasks and has proven effective for feature identification while significantly improving training efficiency \citep{2018MNRAS.479..415A}. Recent studies have shown that using pretrained models trained on diverse galaxy morphologies can improve performance in identifying more complex features, while also reducing the computational cost and training time compared to training from scratch. Zoobot, a model pretrained on diverse galaxy morphologies \citep{2023JOSS....8.5312W}, has been used to identify lopsidedness in spiral galaxies \citep{2025arXiv250519583S}. Machine learning methods are therefore well-suited for studying various galaxy morphologies. Warped structures are typically faint and become prominent in the outer regions of galactic discs, making their detection observationally challenging. Deep learning models can aid in identifying such faint features in galaxy outskirts, while transfer learning enables efficient application of these methods to upcoming deep, wide-field surveys such as Euclid \citep{2025arXiv250315302E} and the Vera Rubin Observatory Legacy Survey of Space and Time (LSST) \citep{2019ApJ...873..111I}, facilitating the identification of larger samples of warped systems and improving our understanding of their origin and evolution.

In this paper, we finetune a deep learning model, Zoobot, to identify warped galaxies from non-warped edge-on systems. The model was trained on edge-on galaxy images from Pan-STARRS, with labels assigned based on warp angles. The trained model was then used to identify a larger sample of warped galaxy candidates from Pan-STARRS. Additionally, we used the model to predict warped galaxies in deeper, higher-resolution imaging from Euclid. We further study the physical properties of warped galaxies and examine the model performance, including interpretability through LayerCAM visualisations. The paper is organised as follows. Section ~\ref{sec:Data} describes the selection criteria for the training sample. Section ~\ref{sec:DCNN} presents the model architecture and training procedure. The results and discussion are presented in Section ~\ref{sec:Results and Discussion}, and finally, Section ~\ref{sec:Conclusion} summarises the main conclusions.

\section{Data}
\label{sec:Data}
The data used in this study are drawn from the catalogue of Edge-on Galaxies in
the Pan-STARRS1 survey (EGIPS) \citep{2022MNRAS.511.3063M}. The Pan-STARRS1 (PS1)
survey is a wide-field optical imaging survey carried out with the 1.8-m
Pan-STARRS telescope at Haleakal\=a Observatory, Hawaii
\citep{2016arXiv161205560C}. Observations are obtained in five broadband filters
($g$, $r$, $i$, $z$, and $y$), covering the entire northern sky and part of the
southern hemisphere down to declinations of $\delta \gtrsim -30^\circ$. The Pan-STARRS1 images are resampled onto a regular grid of
$4^\circ \times 4^\circ$ projection cells, which are further divided into
$10 \times 10$ skycells. Each skycell spans $0.4^\circ \times 0.4^\circ$ with a
pixel scale of 0.25 arcsec per pixel. The EGIPS catalogue,
constructed from Pan-STARRS PS1 Data Release 2, contains 16,551 nearly edge-on
galaxy candidates distributed over approximately three-quarters of the sky. The catalogue provides homogeneous measurements of astrometry,
SExtractor-based photometry \citep{1996A&AS..117..393B}, and non-parametric morphological parameters derived
from the Pan-STARRS images. Their initial sample selection was performed using
an artificial neural network, followed by visual inspection to remove image
artefacts, misclassified objects, and non-edge-on galaxies. 

\subsection{Sample selection}
\label{sec:Sample selection} 
Since disc warps are observed primarily on galaxies with edge-on orientation, we restrict our analysis to highly inclined galaxies, having an inclination of 85-90 degrees. The inclination angles for the galaxies are computed using the semi-major (stdA) and semi-minor (stdB) axes provided by the photometry table collected from the SExtractor data. 
The inclination of the galaxy is calculated adopting \citep{2008MNRAS.387.1099P}. Assuming an intrinsic disc thickness of $q_0 = 0.2$, the
inclination angle $i$ is derived as

\begin{equation}
\cos^2 i = \frac{q^2 - q_0^2}{1 - q_0^2},
\end{equation}
where $q = b/a$ . 

Based on the inclination, we separate the galaxies into three inclination classes as ``Gold" with $i = 90^\circ$ corresponding to a perfectly edge-on galaxy, ``Silver" with $85^\circ \leq i < 90^\circ$, which are classified as nearly edge-on galaxies and ``Other" for galaxies with $ i < 85^\circ$. Additionally, each galaxy in the EGIPS catalogue is associated with classifications from a citizen science survey based on visual inspection. A class is assigned according to the dominant vote fraction, with galaxies categorised as ``Good" when the highest vote fraction exceeds 80\%, ``Acceptable" when it exceeds 50\%, and ``Unsuitable" otherwise. Galaxies lacking sufficient voting information are flagged separately and excluded from the main analysis. In this work, we focus on the highly inclined galaxies with Gold inclination ($i = 90^\circ$) and Silver inclination class $85^\circ \leq i < 90^\circ$. This sample is further subdivided into six classes based on the votes and inclination, ``Good\_Gold", ``Good\_Silver", ``Acceptable\_Gold", ``Acceptable\_Silver", ``Unsuitable\_Gold", and ``Unsuitable\_Silver". This approach produces a consistent sample of edge-on galaxies, suitable for studying stellar disc warps. Additionally, the EGIPS visual classification scheme has been designed to assess image quality and general morphology, but does not incorporate a specific category for warped discs. Consequently, strongly warped edge-on galaxies were in some cases interpreted as spiral or disturbed systems and classified as Unsuitable. Retaining these objects ensures that the analysis is not systematically biased against galaxies with large-amplitude warps. 

\begin{figure}
    \centering
    \includegraphics[width=\columnwidth]{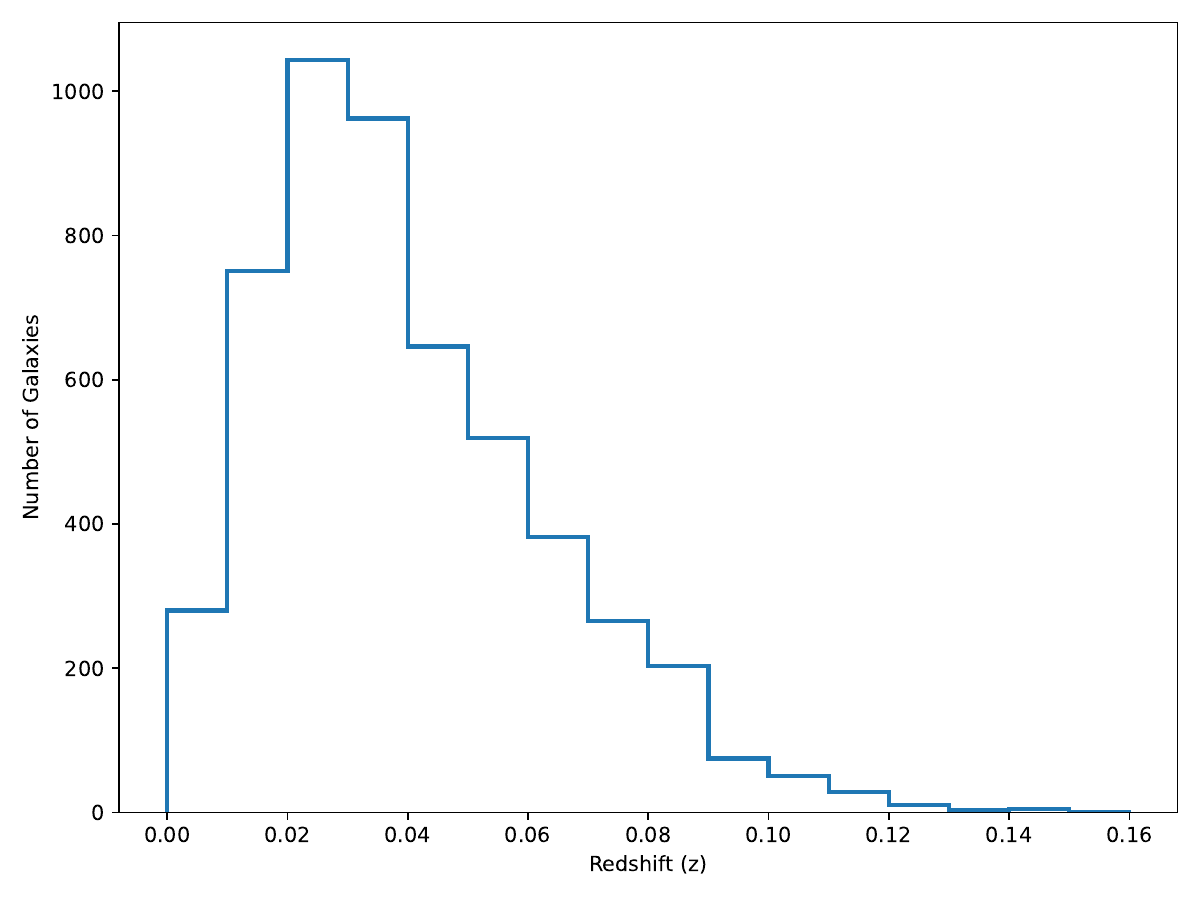}
    \caption{Distribution of redshift for the sample of highly inclined edge-on galaxies ($i \geq 85^\circ$).}
    \label{fig:redshift_dist}
\end{figure}

Visual inspection was performed to minimise projection effects, resulting in a final sample of 8497 highly inclined edge-on galaxies. The redshift distribution of our sample of highly inclined galaxies (inclination $\geq 85^\circ$) is shown in Fig.~\ref{fig:redshift_dist}. The EGIPS catalogue provides redshift measurements for approximately 63\% of the galaxies based on the HyperLeda database \citep{2022MNRAS.511.3063M}. The distribution indicates that our sample is dominated by nearby galaxies. This behaviour is consistent with the parent EGIPS catalogue, which also exhibits a deficit of galaxies at higher redshift. As these samples are based on morphological selection, structural features such as warps become increasingly difficult to detect at larger distances due to resolution limits and low signal-to-noise ratios. The complete dataset used in this work is publicly accessible on GitHub\footnote{\url{https://github.com/saranya-suguna/Warped_DCNN}}. For each galaxy in the selected subsamples, FITS image cutouts of $300 \times 300$ pixels are downloaded from the Pan-STARRS PS1 DR2 archive in the $i$-band, centred on the galaxy coordinates \citep{2016arXiv161205560C}. Given the Pan-STARRS pixel scale of 0.25 arcsec per pixel, this corresponds to a field of view of $75'' \times 75''$ ($1.25 \times 1.25$ arcmin). The $i$-band images are selected as it provides better signal-to-noise imaging of stellar discs while minimising the effects of dust extinction and sky background, making it suitable for morphological studies of edge-on galaxies. However, training with $r$-band images yields similar performance, which is consistent with \citet{2022ApJ...935...48Z}, where no dependence on optical bands was found for warp measurements. Initial training is conducted using DSS images from warped galaxy catalogues in VizieR. We also explore higher redshift HST/JWST images \citep{2025A&A...697L...1R}, which are found to be less suitable for training due
to their reduced contrast in the outer regions of the galaxy. We therefore adopt the Pan-STARRS dataset, which provides a larger and more homogeneous sample.

\subsection{Warp Angle Measurement}
\label{sec:warp_measurement}

To provide quantitative and homogeneous labels for the deep convolutional neural network (DCNN), we measure the degree of warping in each galaxy. Warp angles in edge-on galaxies have been quantified using various techniques. Earlier studies have relied on visual inspection to identify and classify warps, while more quantitative approaches include isophotal fitting \citep{2006NewA...11..293A, 2016MNRAS.461.4233R}, and Fourier or bending-mode decomposition \citep{1988MNRAS.234..873S}. Kinematic approaches, such as tilted-ring modelling, have also been widely used to characterise warps in HI discs \citep{1990ApJ...352...15B, 2002A&A...394..769G}. Warps trace the vertical displacement of the disc as a function of radius \citep{2002A&A...391..519C, 2001A&A...373..402S}. More recent works employ automated techniques to quantify galactic warps. \citet{2016MNRAS.461.4233R} uses isophotal maps to identify warps from position angle variations between the inner and outer disc, while \citet{2022ApJ...935...48Z} adopts a Gaussian disc tracing approach, enabling scalable analysis for large samples. We experiment with two-dimensional ellipse fitting to outer isophotes, along with background subtraction, aggressive masking of foreground and background sources. However, this approach prove unstable in low surface-brightness outer discs and in galaxies with asymmetric light distributions. Since Pan-STARRS images are not explicitly background-subtracted and can contain variable noise and contamination from nearby sources, these methods did not yield reliable measurements across the full sample. Given these constraints, we find that a Gaussian-weighted centroid method for tracing the disc mid-plane, as described by \citet{2022ApJ...935...48Z}. It provides the most stable and reproducible measurements for our sample.

Each galaxy is rotated to align its major axis horizontally using the position angle (PA), such that the disc mid-plane lies parallel to the image $x$-axis. After geometric alignment, images are cropped using a rectangular aperture defined by the semi-major and semi-minor axes (StdA and StdB), converted to pixel units using the Pan-STARRS pixel scale (0.5 arc seconds). The resulting rotated and cropped images contain the full radial extent of the galactic disc with minimal background contamination. Each image is first smoothed with a Gaussian kernel ($\sigma = 1.5$ pixels) to suppress small-scale noise. The disc is then sampled along the $x$-axis using radial bins of fixed width (5 pixels). For each radial bin, the vertical intensity profile is constructed by averaging the flux within the bin. The vertical position of the disc mid-plane is determined via a Gaussian-weighted centroid of the brightest portion of the profile, including pixels exceeding 15\% of the peak normalised intensity. Radial bins with peak flux below $1.8\sigma$ above the local sky background are discarded to avoid noise-dominated regions. This procedure yields a discrete set of mid-plane points $y(x)$ that trace the disc structure across the radius. Outliers are removed using a median absolute deviation (MAD) filter, and the resulting spine was smoothed using a third-order Savitzky-Golay filter to suppress small-scale fluctuations while preserving large-scale bending.

Warps are predominantly formed in the outer disc beyond the bright stellar component \citet{1990ApJ...352...15B}. We define the inner-disc boundaries between $0.3$–$0.5\,\mathrm{StdA}$ and outer-disc boundaries between $0.7$–$0.8\,\mathrm{StdA}$, and found the resulting warp angles to be stable within the estimated measurement uncertainties. We therefore adopt $0.4\,\mathrm{StdA}$ as the inner reference region and $0.75\,\mathrm{StdA}$ as the outer warp region. The projected galactocentric radius is defined as $R = |x|$. The inner disc ($|x| < 0.4\,\mathrm{StdA}$) is assumed to be approximately planar and used to define a reference mid-plane via a linear fit. Warp measurements were performed independently on the left ($x<0$) and right ($x>0$) outer disc regions ($|x| > 0.75\,\mathrm{StdA}$).
For each side, the warp angle is defined as
\begin{equation}
\psi = \tan^{-1}\left( \frac{d y}{d R} \right)
\end{equation}
where $dy/dR$ is the slope of a linear fit to the outer-disc spine points expressed in projected galactocentric radius $R$. 

\begin{figure*}
    \centering
    
    \begin{minipage}{0.49\textwidth}
        \centering
        \includegraphics[width=\textwidth]{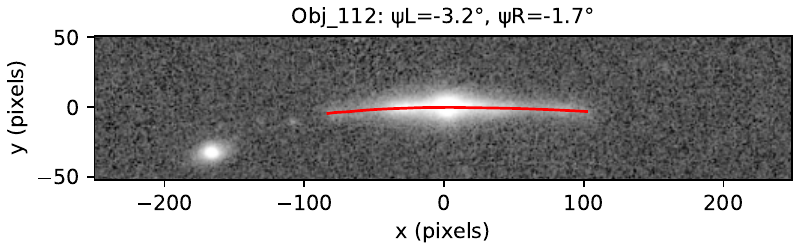}
        {\small (a)}
    \end{minipage}
    \hfill
    \begin{minipage}{0.49\textwidth}
        \centering
        \includegraphics[width=\textwidth]{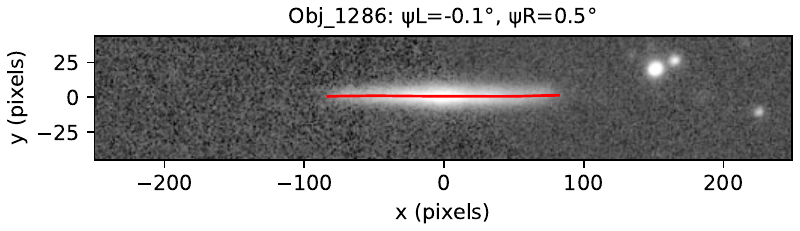}
        {\small (b)}
    \end{minipage}
    
    \caption{Warp angle measurement for an edge-on galaxy. The red curve traces the extracted disc mid-plane (spine) obtained using the Gaussian-weighted centroid method. The inner disc defines a reference plane, while systematic deviations of the outer disc on the left and right sides are used to compute the warp angles $\psi_{\mathrm{L}}$ and $\psi_{\mathrm{R}}$. Left (a) shows a warped galaxy, whereas right (b) shows a non-warped galaxy.}
    
    \label{fig:warp_measurement}
\end{figure*}

Uncertainties in the warp angles are estimated from the scatter of the outer-disc spine points about the corresponding linear fit. The vertical residuals are quantified using the median absolute deviation (MAD), converted to an equivalent Gaussian standard deviation, and propagated into angular units via $\tan^{-1}(\sigma_y / R)$, where $R$ is the median projected radius of the fitted outer region. To quantify the overall warp strength of each galaxy, we define a scalar warp score as the maximum absolute value of the left and right warp angles. Based on this score, galaxies were classified into non-warped ($\psi < 0.5^\circ$) and warped ($\psi \geq 0.5^\circ$) galaxies. An example of the disc spine extraction and warp angle definition is shown in Fig.~\ref{fig:warp_measurement}, where panel (a) corresponds to a warped galaxy and panel (b) to a non-warped galaxy. The figure illustrates how the spine traces the galactic disc. However, Pan-STARRS images are not fully background-subtracted and can contain residual sky features, noise, and nearby sources, which can introduce uncertainties in the measured warp angles, especially in the faint outer parts of the disc. Therefore, visual inspection is subsequently performed to verify and refine the automated classifications before constructing the final training sample. The final CNN training dataset consists of 1000 edge-on galaxies, which include 500 warped and 500 non-warped systems.
The remaining images are kept for testing of the trained model.

\section{Deep Convolutional Neural Network Classification}
\label{sec:DCNN}

To train and classify edge-on galaxies as warped or non-warped, we employ a convolutional neural network based on the Zoobot framework \citep{2023JOSS....8.5312W}. Zoobot is a deep-learning architecture developed for galaxy morphology classification and is pretrained on large datasets of Galaxy Zoo-labelled images, enabling it to learn general morphological representations from diverse galaxy populations. In this work, we adopt a transfer learning approach. We initialise the ConvNeXT-Nano encoder with pretrained Zoobot weights and finetune the higher layers for our binary warp classification task, while retaining the lower-level feature representations learned from large-scale morphology datasets. A custom classification head is appended to adapt the model to the two-class problem. This strategy allows the network to leverage previously learned structural features while specialising in identifying warp signatures in edge-on galaxies.

\subsection{Input data}

The network is trained using $i$-band Pan-STARRS1 FITS image cutouts, with their original orientations. The images are processed using the Zoobot FITS pre-processing pipeline. Pixel values are rescaled using an arcsinh transformation and clipped using percentile-based limits (0–99.7 \%) to suppress extreme outliers while retaining faint disc structure. All images are resized to $224 \times 224$ pixels. The dataset is randomly divided into training (60\%), validation (20\%), and test (20\%) subsets using a fixed random seed. The classification labels are derived from quantitative warp angle measurements (Section~\ref{sec:warp_measurement}) and refined through visual inspection. Augmentation on the training images is kept minimal, with random horizontal flips applied with a probability of 0.5, while rotations were not used. 

\subsection{Training Zoobot}
The CNN architecture consists of a pretrained Zoobot encoder based on the ConvNeXT-Nano backbone, followed by a custom classification head. The encoder is initialised with weights trained on various galaxy morphologies and used as a feature-extraction configuration, returning multi-scale feature maps. The final encoder feature map is passed to a classification head comprising a global average pooling layer, a fully connected layer with 128 units, a ReLU activation, a dropout layer with a rate of 0.2, and a final linear layer producing two output logits corresponding to warped and non-warped classes.

Training is performed using the AdamW optimiser \citep{2014arXiv1412.6980K} with a learning rate of $10^{-5}$ and weight decay of 0.05. We adopt a layer-wise learning rate decay strategy to enable stable finetuning of the pretrained encoder. The classification head is trained with the full learning rate, while consecutive deeper stages are trained with reduced learning rates scaled by a decay factor of 0.5 per stage. Earlier encoder layers remain frozen. This approach allows for adaptation to warp-specific morphology while preserving low-level feature representations learned from large galaxy datasets. The network is trained for 50 epochs with a batch size of 64. Early stopping is implemented based on the validation loss with a patience of five epochs to prevent overfitting. The model parameters corresponding to the lowest validation loss are retained as the final model. The model with the lowest validation loss is used for prediction on the test set. The softmax activation function is used to convert the class score into class probabilities. To introduce randomisation, we repeat the training for 10 independent runs. For each run, the dataset is reshuffled with a fixed random seed before splitting into training, validation, and test subsets. This procedure reduces sensitivity to sampling variance and ensures that different galaxies contribute to the test set across runs. All models are trained and evaluated using PyTorch \citet{2019arXiv191201703P}.

\section{Results and Discussion}
\label{sec:Results and Discussion}
\subsection{Model interpretability}

To ensure that the CNN model has learned the features rather than background noise or artefacts, we employ Explainable Artificial Intelligence techniques. These tools highlight the regions that contribute most strongly to a given prediction. One widely used method is Gradient-weighted Class Activation Mapping (Grad-CAM) \citep{2016arXiv161002391S}, which utilises gradients flowing into convolutional layers to generate class-discriminative localisation maps. An extension of this approach, LayerCam, improves spatial localisation by preserving pixel-wise gradient information within intermediate feature maps \citep{Jiang2021LayerCAM}.

\begin{figure}
    \centering    
    \includegraphics[width=\columnwidth]{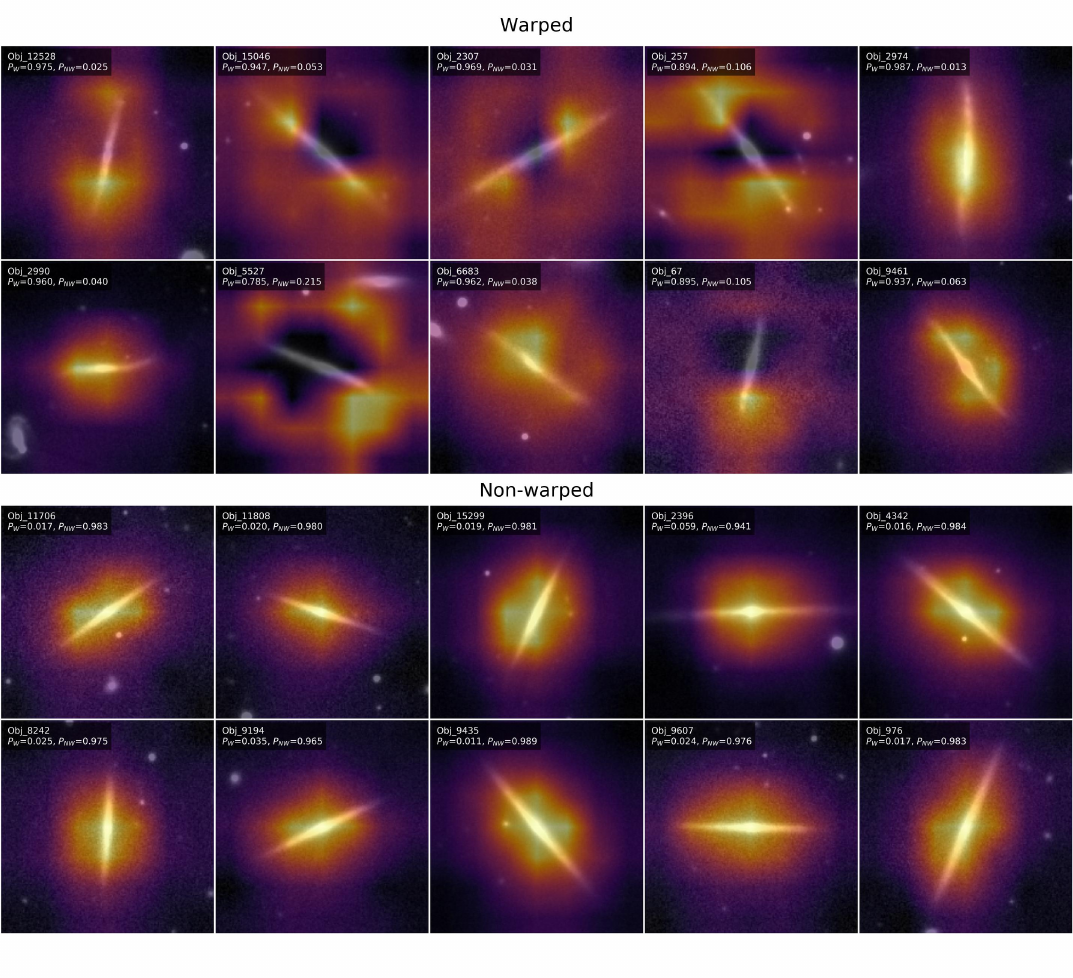}
    \caption{LayerCam visualisations for representative Warped galaxies (top) and Non-warped galaxies (bottom) galaxies. Warped systems show enhanced activation in the outer disc regions where the mid-plane deviates from linearity, while non-warped systems exhibit symmetric activation distributed along the disc plane.}
    \label{fig:layercam_comparison}
\end{figure}

LayerCam applies a gradient-based class activation mapping method to the trained classifier to interpret its predictions. For each test galaxy, a forward pass is performed to obtain class probabilities. The activation map is then computed with respect to the predicted class using gradients from a deep convolutional layer. The resulting map is upsampled to the input image resolution, normalised, and overlaid on the corresponding flux-scaled FITS image. These activation maps are used to assess whether the classifier focuses on physically relevant morphological features, such as warped disc structures or symmetric edge-on discs.
Fig.~\ref{fig:layercam_comparison} shows the LayerCAM visualisations for a sample of warped and non-warped galaxies in the test set. The heat maps highlight the spatial regions that contribute the most strongly to the model’s classification decision. The model primarily focuses on the main galaxy even in the presence of background sources, indicating that the classification is driven by morphological features.

For the warped galaxies, the heatmaps are concentrated toward the outer disc regions, where the stellar mid-plane deviates from linearity. The enhanced response in these regions indicates that the network is primarily sensitive to morphological asymmetries in the outer disc structure, consistent with the physical definition of galactic warps. Notably, the central bulge and inner disc contribute comparatively less to the classification, suggesting that the model does not rely on global brightness concentration but instead focuses on structural deviations at larger radii. In contrast, for the non-warped edge-on galaxies, the LayerCam maps exhibit a more symmetric and centrally distributed activation pattern along the disc plane. This symmetric distribution reflects the model’s recognition of a coherent, undistorted disc morphology, where the vertical structure remains approximately constant across galactocentric radius. The distinction between asymmetric outer-disc activations in warped systems and symmetric mid-plane activations in non-warped systems demonstrates that the convolutional network is not making spurious classifications based on noise or background artefacts. Instead, it appears to capture physically meaningful morphological features associated with disc warping. Therefore, these visualisations provide interpretability and confirm that the classifier’s decision-making process is consistent with established astrophysical definitions of warped and non-warped edge-on galaxies.

\subsection{Model evaluation}
\begin{figure}
    \centering
    \includegraphics[width=\columnwidth]{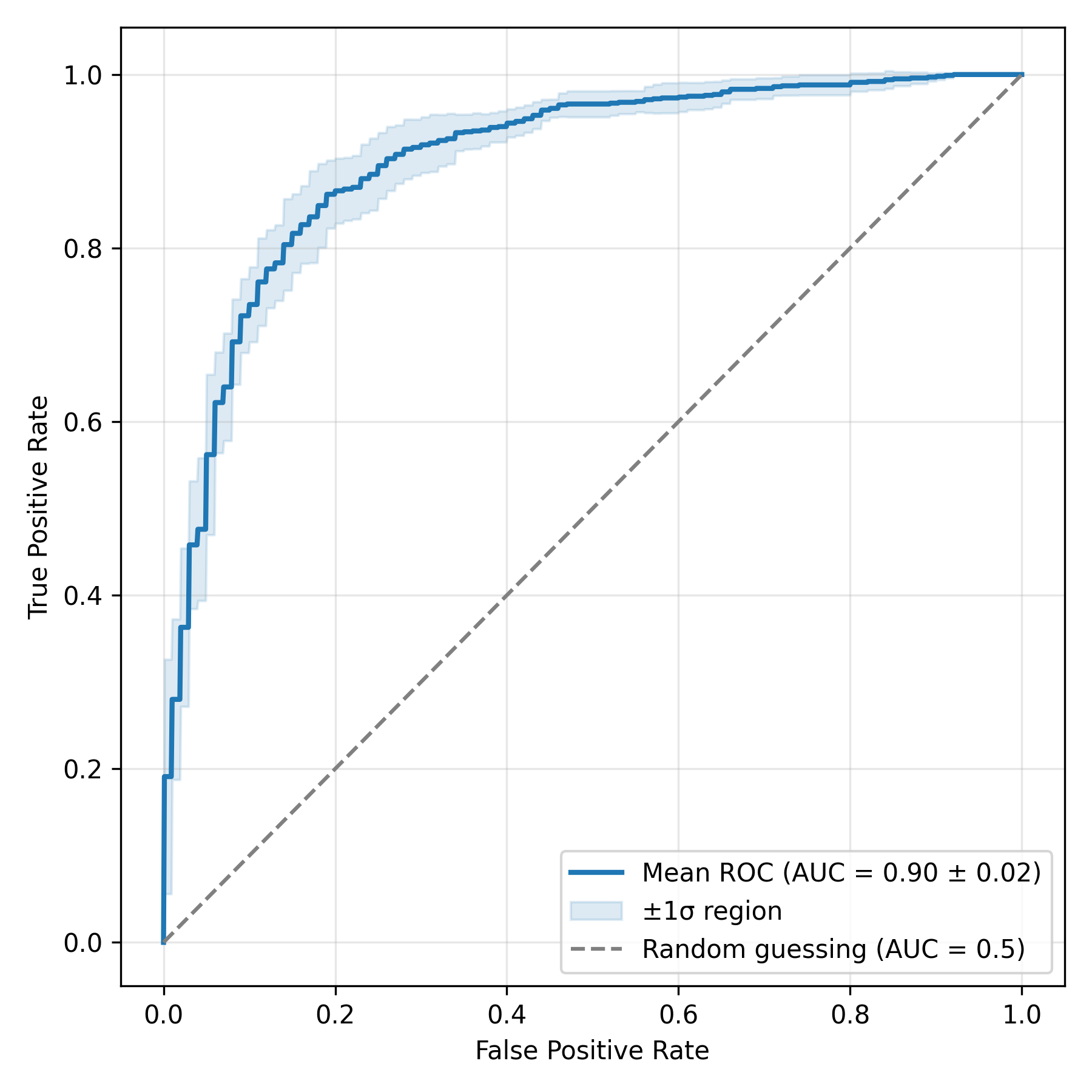}
    \caption{Mean receiver operating characteristic curve across ten independent runs. The shaded region shows the $1\sigma$ spread, indicating the variability in performance due to dataset splits and stochastic optimisation.}
    \label{fig:roc}
\end{figure}

To assess the performance and stability of the warp classifier, we conduct a comprehensive evaluation across 10 independent training runs. For each run, the model saved at the epoch with the lowest validation loss was selected for testing. We computed standard classification metrics, including accuracy, precision, recall, F\(_1\)-score, and the receiver operating characteristic (ROC) curve with its corresponding area under the curve (AUC), which provides a measure of the classifier’s discriminative ability independent of any decision threshold. The mean ROC curve obtained from ten independent runs is shown in Fig.~\ref{fig:roc}. The curve rises steeply at low false-positive rates, reaching high true-positive rates even at modest thresholds, indicating strong separability between warped and non-warped galaxies. The mean AUC of $0.90 \pm 0.02$ shows consistent performance across runs, and the narrow $1\sigma$ range indicates that the results are not strongly affected by data splits or random initialisation. Although the overall classification accuracy is $83 \pm 2.35\%$, the higher AUC indicates that the network reliably ranks galaxies by warp likelihood even when a fixed decision threshold introduces misclassifications near the class boundary. 

\begin{figure}
    \centering
    \includegraphics[width=\columnwidth]{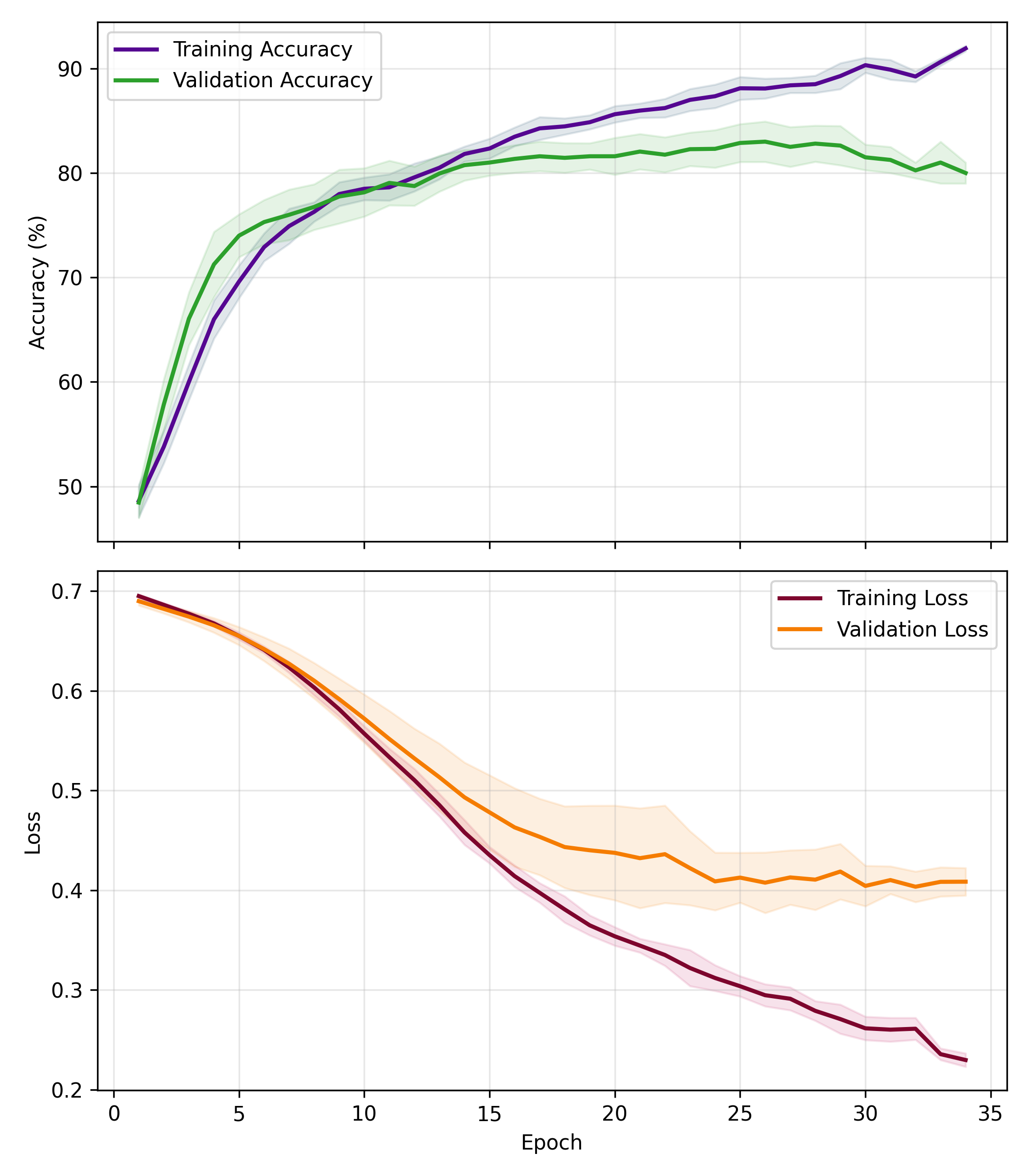}
    \caption{Epoch-wise accuracy (top) and loss (bottom) averaged across ten independent runs. Shaded regions indicate the $1\sigma$ spread due to run-to-run variability.}
    \label{fig:loss_acc}
\end{figure}

\begin{figure}
    \centering
    \includegraphics[width=\columnwidth]{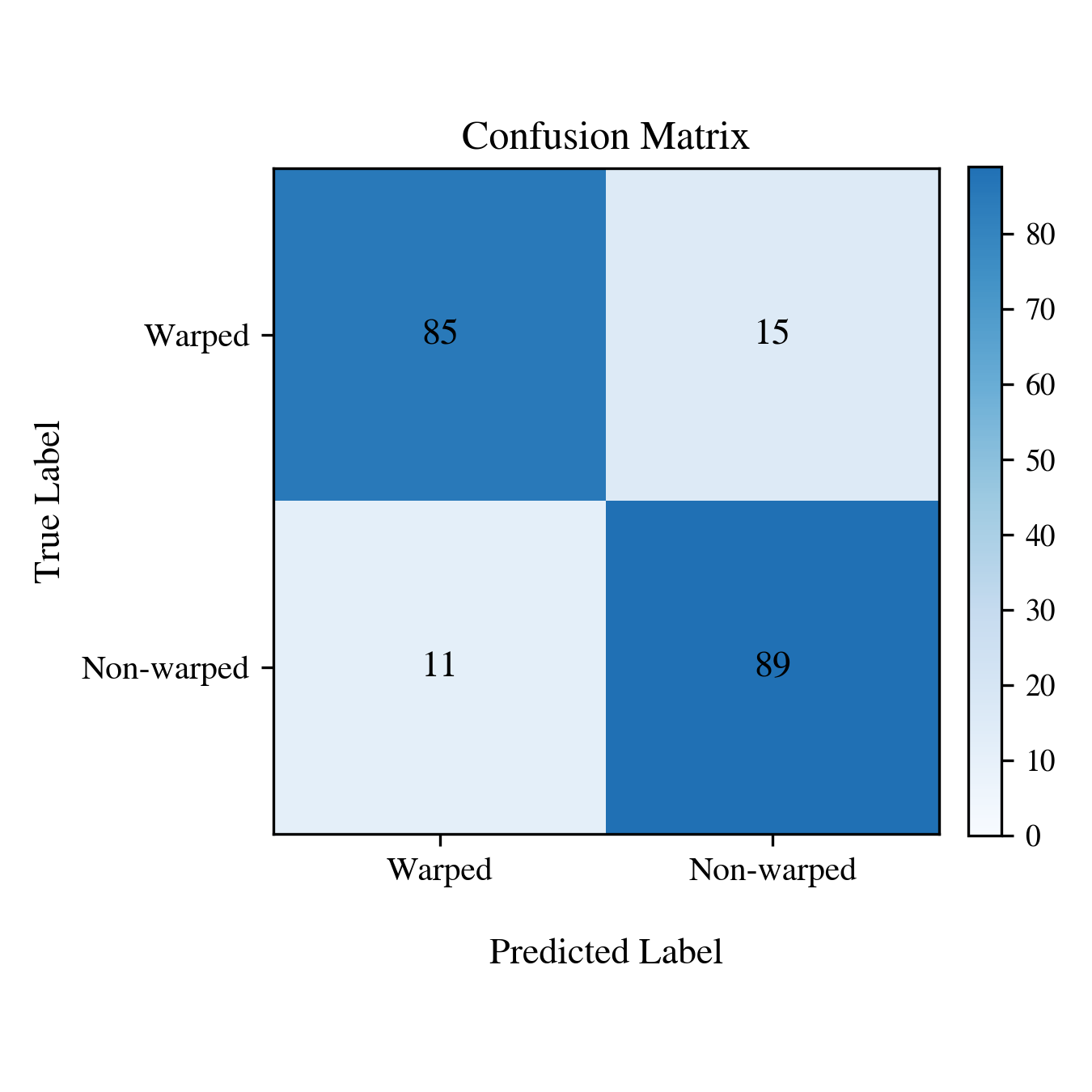}
    \caption{Confusion matrix for the best-performing model evaluated on the test set. }
    \label{fig:confmat}
\end{figure}

The training and validation accuracy and loss curves averaged over all runs to evaluate convergence behaviour in Fig.~\ref{fig:loss_acc}. Training accuracy increases steadily to $\sim92$ \%, while validation accuracy saturates near $\sim80$--$83$ \% after approximately 15 epochs. A modest yet consistent gap between training and validation performance suggests mild overfitting at later epochs, which justifies the use of early stopping. Similarly, the training loss decreases monotonically, whereas the validation loss flattens after $\sim15$--$20$ epochs and stabilises around $\sim0.4$. The absence of divergence or sharp oscillations indicates stable optimisation. The plateau in validation performance suggests that the model approaches an intrinsic limit imposed by the data rather than optimisation instability. The confusion matrix for the best-performing model (selected by the lowest validation loss) is shown in Fig.~\ref{fig:confmat}. This provides a granular view of true and false predictions for each class on the test set.

The summary statistics in Table~\ref{tab:train_test_prediction_table} report mean precision, recall, and F\(_1\)-score for each class across all runs, showing balanced performance for warped and non-warped galaxies. The balanced precision and recall for both warped and non-warped classes (Table~\ref{tab:train_test_prediction_table}) further show that the classifier does not preferentially favour one morphology over the other. The classification is limited by observational constraints and intrinsic morphological complexity. Warp signatures are often subtle and are most prominent in the faint outer disc regions, where the signal-to-noise ratio in edge-on galaxy images can be relatively low. Small asymmetries, projection effects, and background fluctuations further introduce ambiguity in morphological classification. The trained model achieves an accuracy of $\sim83\%$ on the test sample with an AUC of 0.90 for 10 independent trials.
\begin{table}
\centering
\caption{Mean classification metrics across ten independent runs. Values are reported as mean $\pm$ standard deviation.}
\label{tab:train_test_prediction_table}
\begin{tabular}{cccc}
\hline
Class & Precision & Recall & F1-score \\
\hline
Warped & 0.84 $\pm$ 0.03 & 0.82 $\pm$ 0.03 & 0.83 $\pm$ 0.02 \\
Non-warped & 0.83 $\pm$ 0.02 & 0.84 $\pm$ 0.04 & 0.83 $\pm$ 0.02 \\
 &  &  &  \\
 & AUC & 0.90 $\pm$ 0.02 &  \\
 & Accuracy & 83 $\pm$ 2.24 \% &  \\
\hline
\end{tabular}
\end{table}

As discussed by \citet{2025arXiv250315310E}, deep learning does not overcome the fundamental limitations of visual galaxy morphology, but instead scales human classification to larger datasets. Since human responses are used as ground truth, the labels reflect observable features rather than intrinsic galaxy structure, and observational effects such as redshift, angular size, and image quality remain embedded in both the data and model predictions. The primary limitation, therefore, arises from the information content of the images. Future progress in large-scale morphology studies is expected to depend not only on improved models but also on combining machine learning with simulations and statistical methods to better account for observational biases and recover intrinsic galaxy properties.

\subsection{Prediction on Pan-STARRS images}

\begin{figure*}
    \centering

    % ---- (a) ----
    \textbf{(a) Pan-STARRS: sample of predicted warped galaxies}
    
    \includegraphics[width=0.8\textwidth]{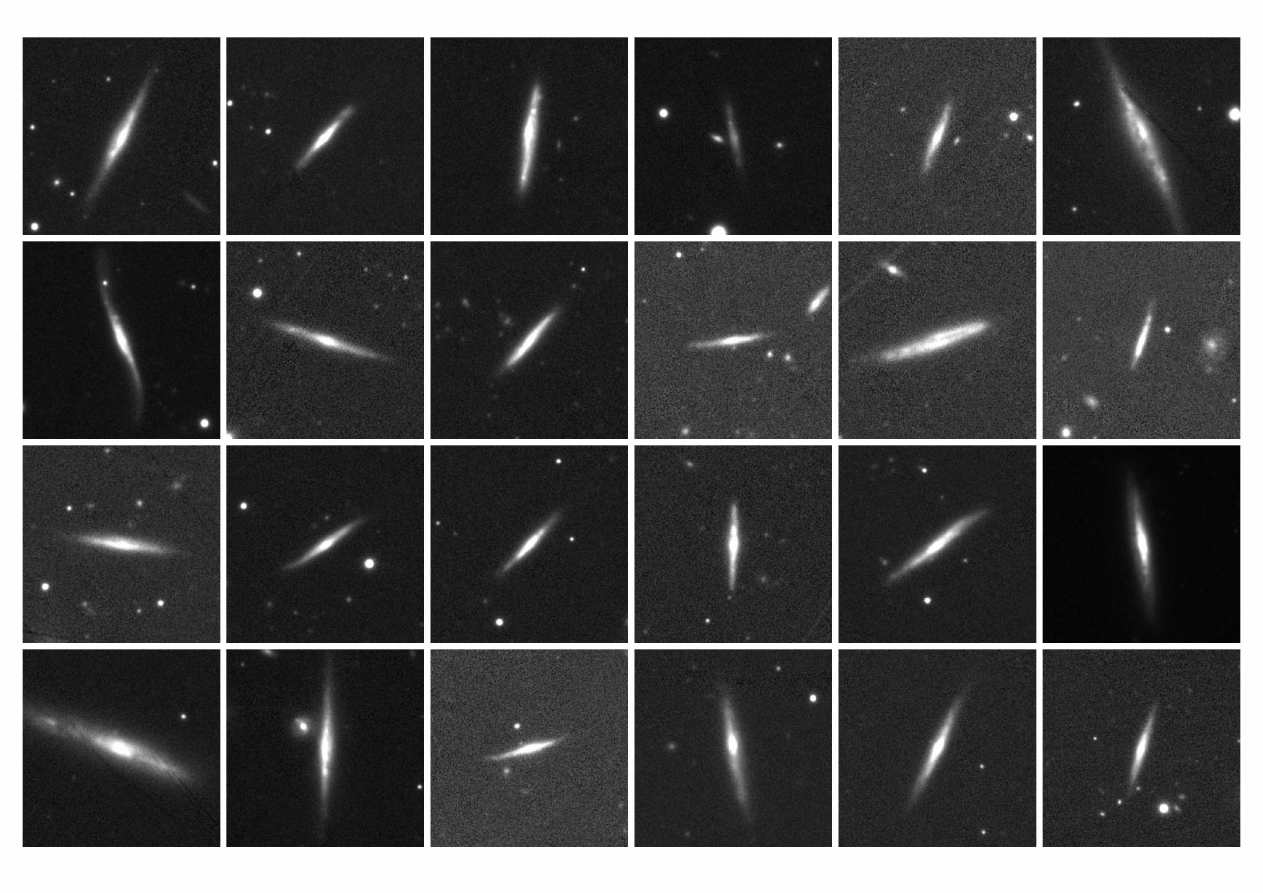}
    
    % ---- (b) ----
    \textbf{(b) Pan-STARRS: sample of predicted non-warped galaxies}
    
    \includegraphics[width=0.8\textwidth]{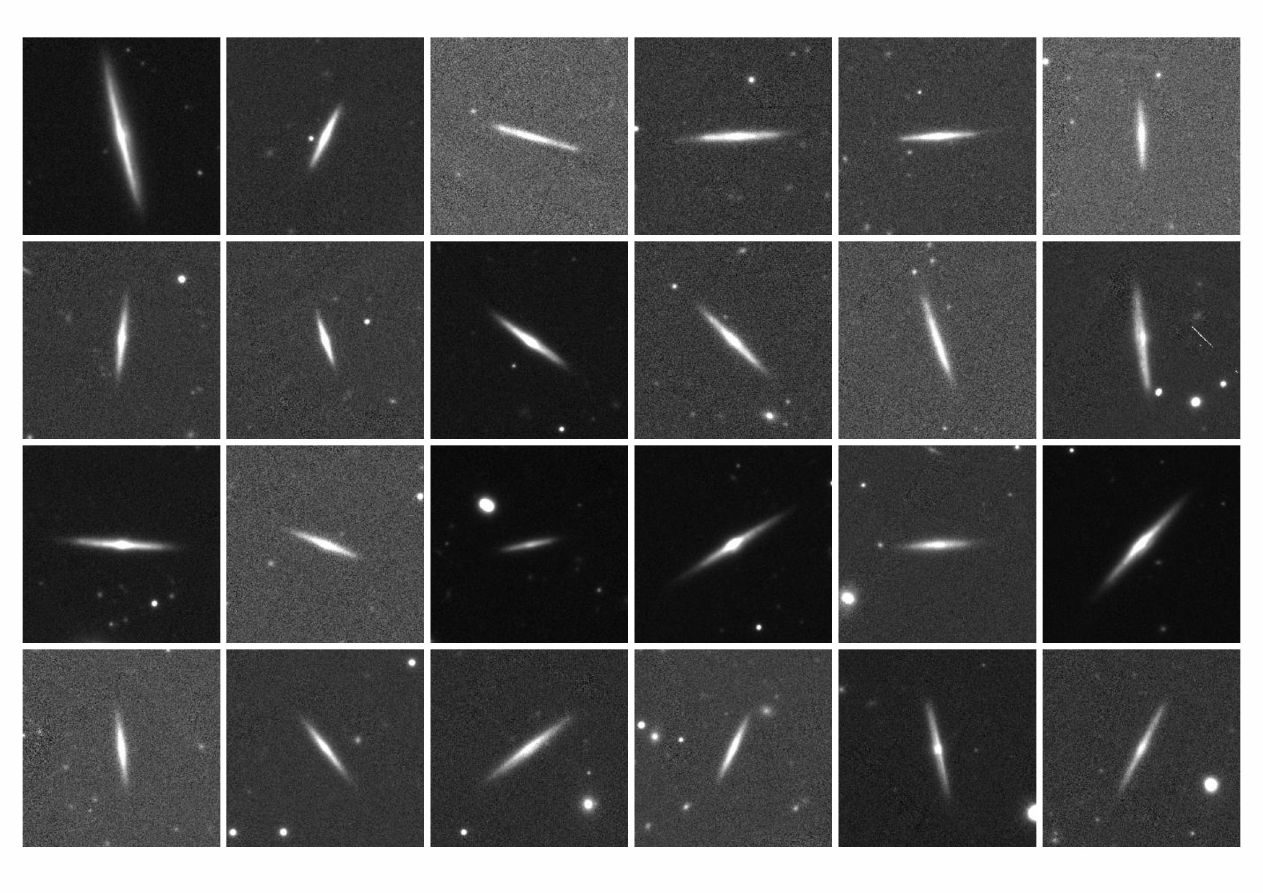}

    \caption{A subset of high confidence predictions on the Pan-STARRS EGIPS catalogue. 
    (a) Warped galaxies. (b) Non-warped galaxies.}
    
    \label{fig:pan_highconf}
\end{figure*}

The best-performing model was subsequently applied to a larger sample of 5393 edge-on galaxies. In addition, we extend the analysis to include nearly edge-on galaxies from the Silver inclination class $85^\circ \leq i < 90^\circ$. After careful visual inspection to minimise projection effects, this adds 2104 galaxies to the sample, resulting in a total test set to 8497 galaxies. To construct a high confidence subset, we select galaxies with prediction probabilities above 85\%. This results in 3486 high-confidence classifications, of which 2088 are identified as warped and 1398 as non-warped. A summary of the prediction probabilities and the corresponding number of images in each class for the Pan-STARRS prediction sample is given in Table~\ref{tab:pan_prediction_summary}. A subset of high-confidence warped and non-warped galaxies from the Pan-STARRS EGIPS test sample is shown in Figure~\ref{fig:pan_highconf}
\begin{table}
\centering
\caption{Summary of the prediction probabilities and the corresponding number of images in each class for Pan-STARRS prediction sample.}
\label{tab:pan_prediction_summary}
\begin{tabular}{ccc}
\hline
Prediction Probability & Unwarped & Warped \\
\hline
90\% and above & 1306 & 2409 \\
85-90 \% & 339 & 512 \\
85\% and below & 1331 & 1577 \\
\hline
\end{tabular}
\end{table}. The full details of the model prediction results are publicly available through our GitHub repository\footnote{\url{https://github.com/saranya-suguna/Warped_DCNN}}.

%----------------------- Panstarrs prediction -------------------------

%----------------Euclid prediction-----------------------------
\begin{figure*}
    \centering

    % ---- (a) ----
    \textbf{(a) Euclid Q1: sample of predicted warped galaxies}
    
    \includegraphics[width=0.8\textwidth]{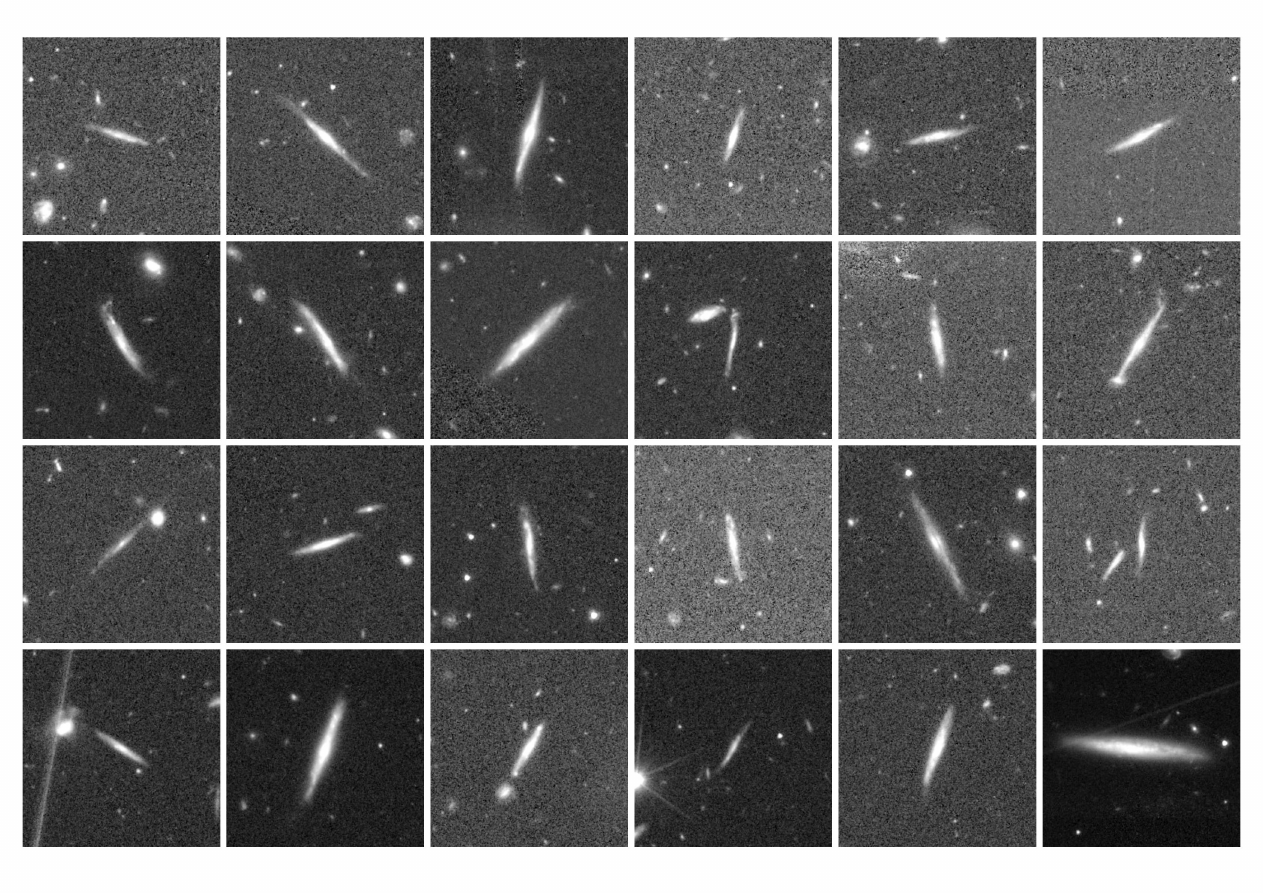}
    
    % ---- (b) ----
    \textbf{(b) Euclid Q1: sample of predicted non-warped galaxies}
    
    \includegraphics[width=0.8\textwidth]{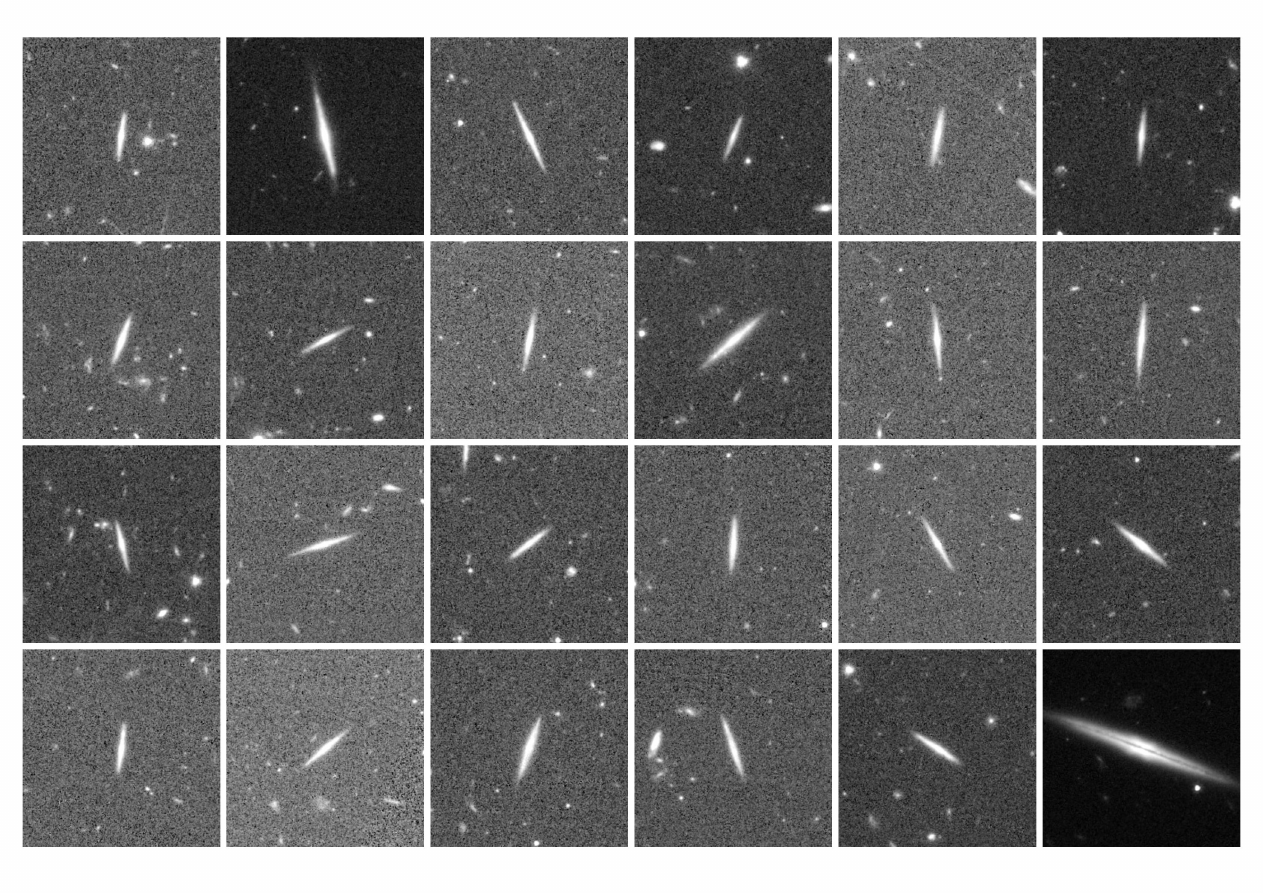}

    \caption{A subset of high confidence predictions on the Euclid Q1 catalogue. 
    (a) Warped galaxies. (b) Non-warped galaxies.}
    
    \label{fig:euclid_highconf}
\end{figure*}
%-------------------------------------------------------------

%--------------------------------------------------------
\begin{figure}
    \centering
    \includegraphics[width=\columnwidth]{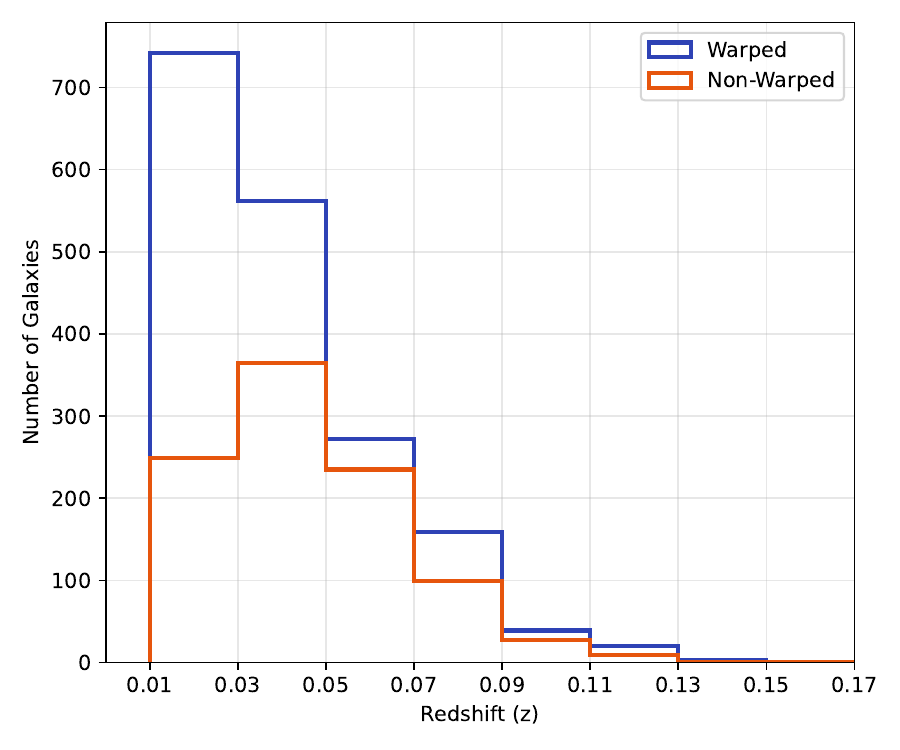}
    \caption{Redshift distribution of warped and non-warped galaxies from the Pan-STARRS EGIPS sample. An imbalance between the two populations is evident in the low-redshift regime (0.01 $\leq$ z $<$ 0.05), while the higher-redshift range is comparatively well sampled.}
    \label{fig:pan_z_dist}
\end{figure}

\begin{figure*}
    \centering
    \includegraphics[width=0.32\textwidth]{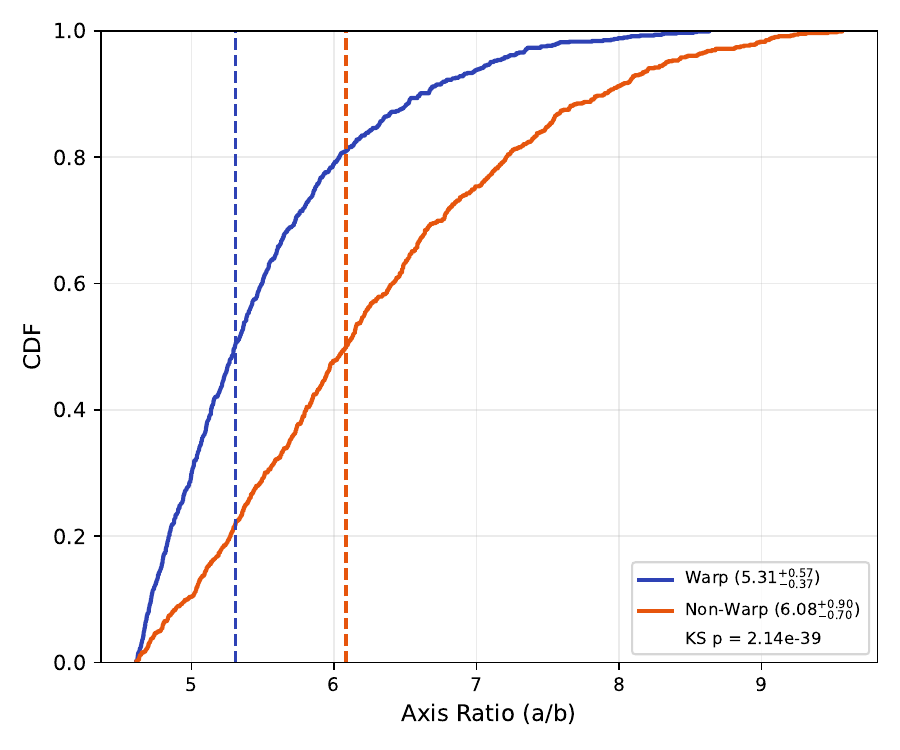}
    \includegraphics[width=0.32\textwidth]{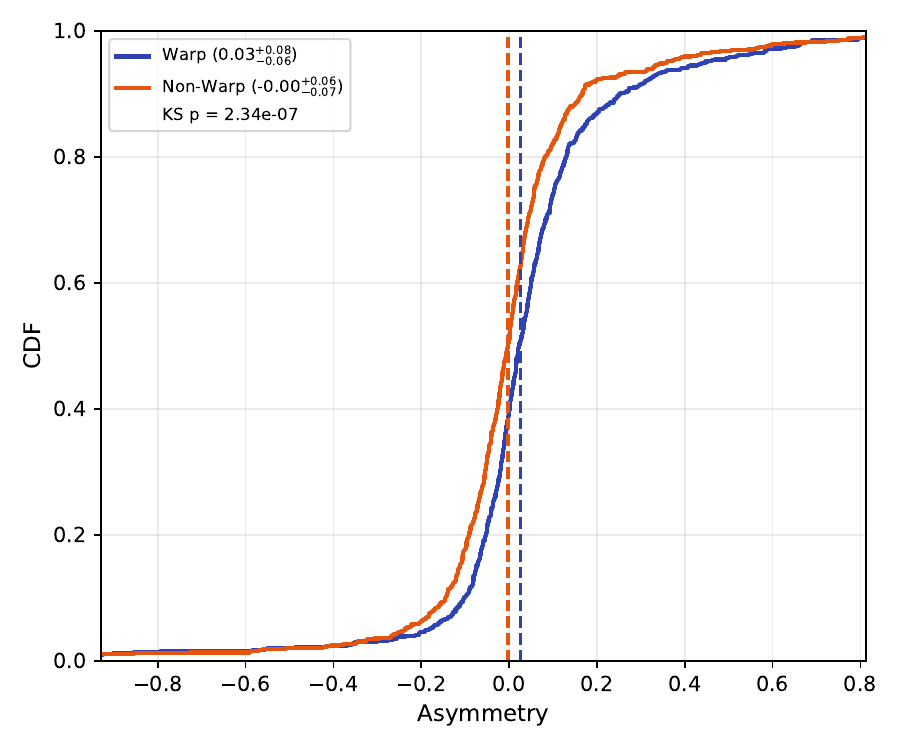}
    \includegraphics[width=0.32\textwidth]{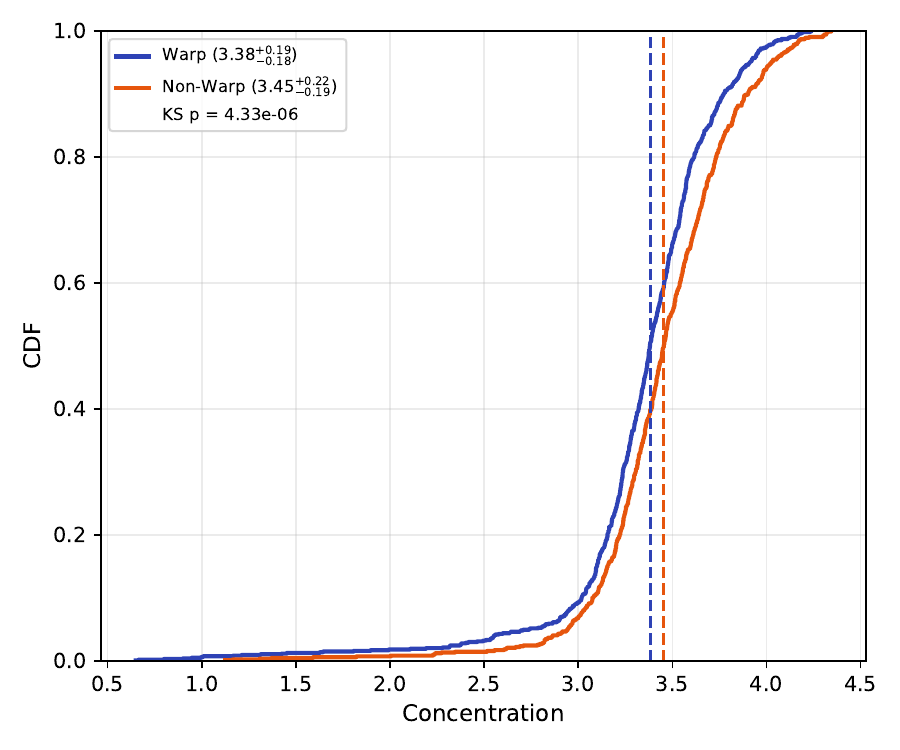}
    \caption{From left to right: Cumulative distribution functions of axis ratio ($a/b$), asymmetry, and concentration for warped and non-warped galaxies in the Pan-STARRS EGIPS sample. The dashed vertical lines indicate the median values of each population, and the KS $p$-values quantify the statistical significance of the differences.}
    \label{fig:pan_morph}
\end{figure*}

%--------------------------------------------------------------------
\begin{figure*}
    \centering
    \includegraphics[width=0.32\textwidth]{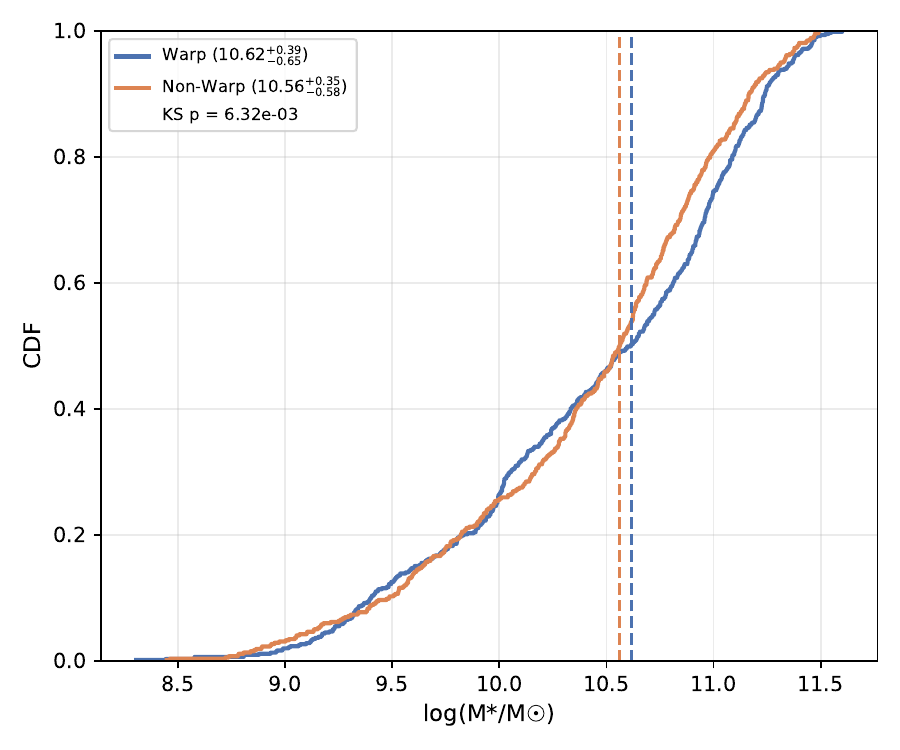}
    \includegraphics[width=0.32\textwidth]{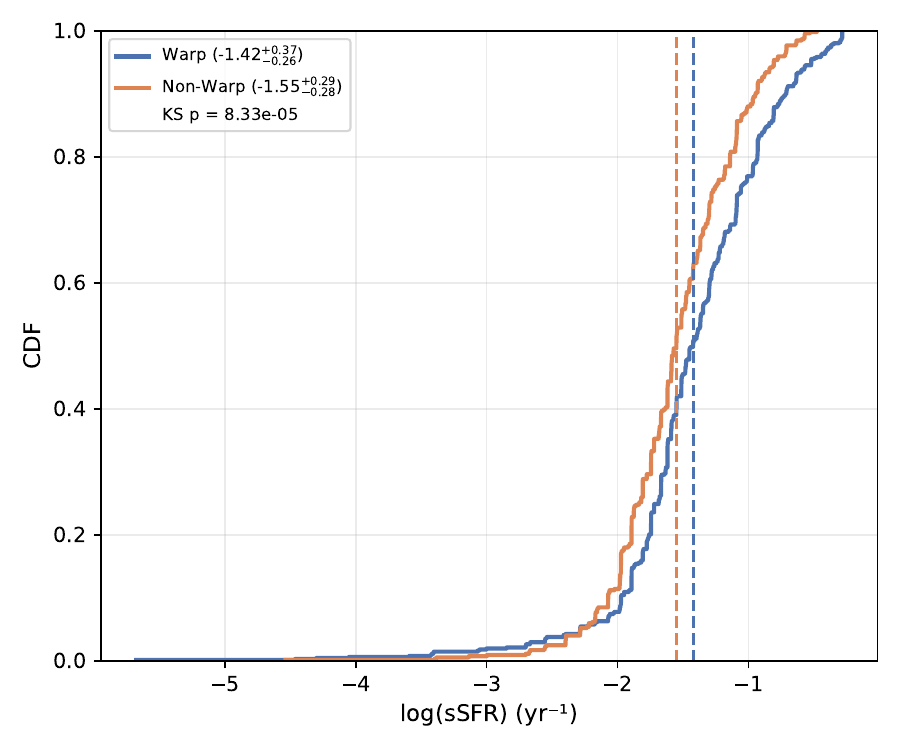}
    \includegraphics[width=0.32\textwidth]{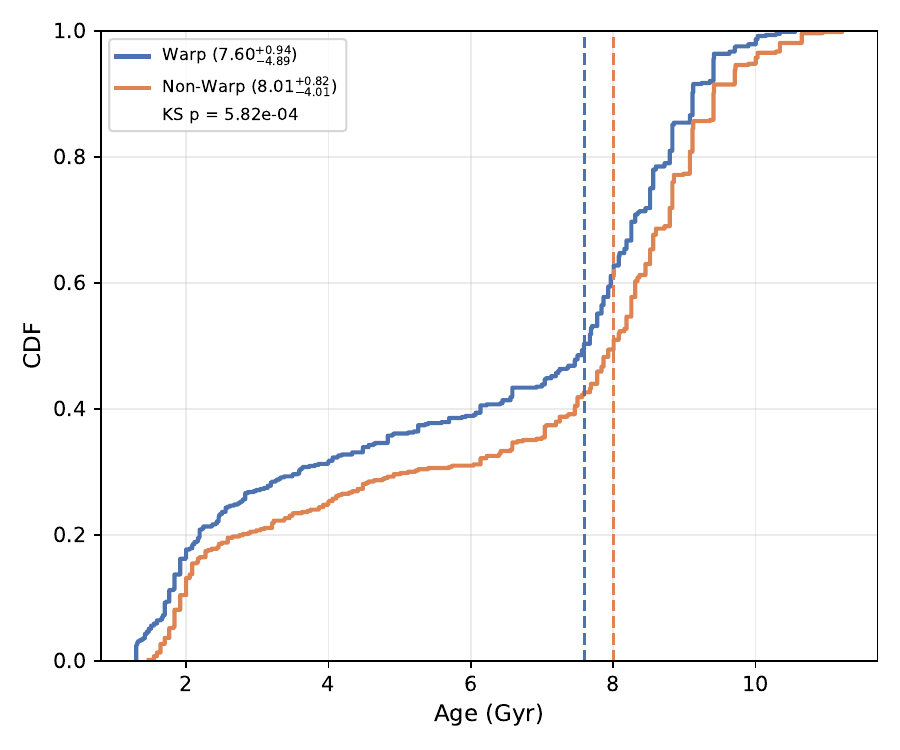}
    \caption{Cumulative distribution functions of physical properties from SDSS for warped and non-warped galaxies. The dashed vertical lines indicate the median values of each population, and the KS $p$-values quantify the statistical significance of the differences. From left to right: stellar mass ($\log(M_\star/M_\odot)$), specific star formation rate (sSFR), and stellar age.}
    \label{fig:pan_sdss}
\end{figure*}

%-------------------------------------------------------------------
\begin{figure*}
    \centering
    \includegraphics[width=0.32\textwidth]{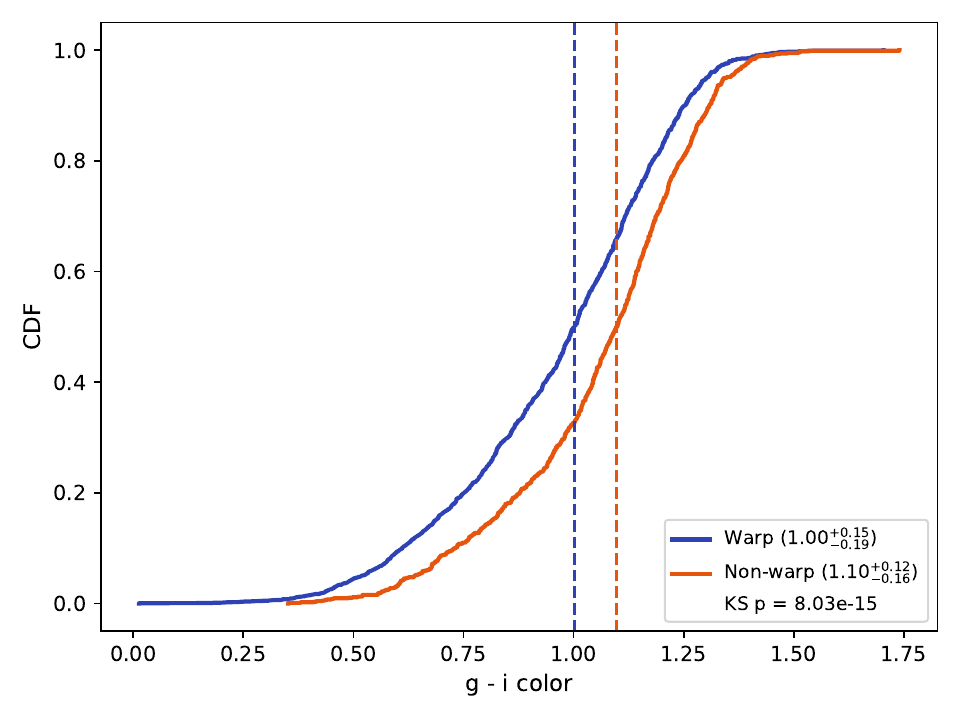}
    \includegraphics[width=0.32\textwidth]{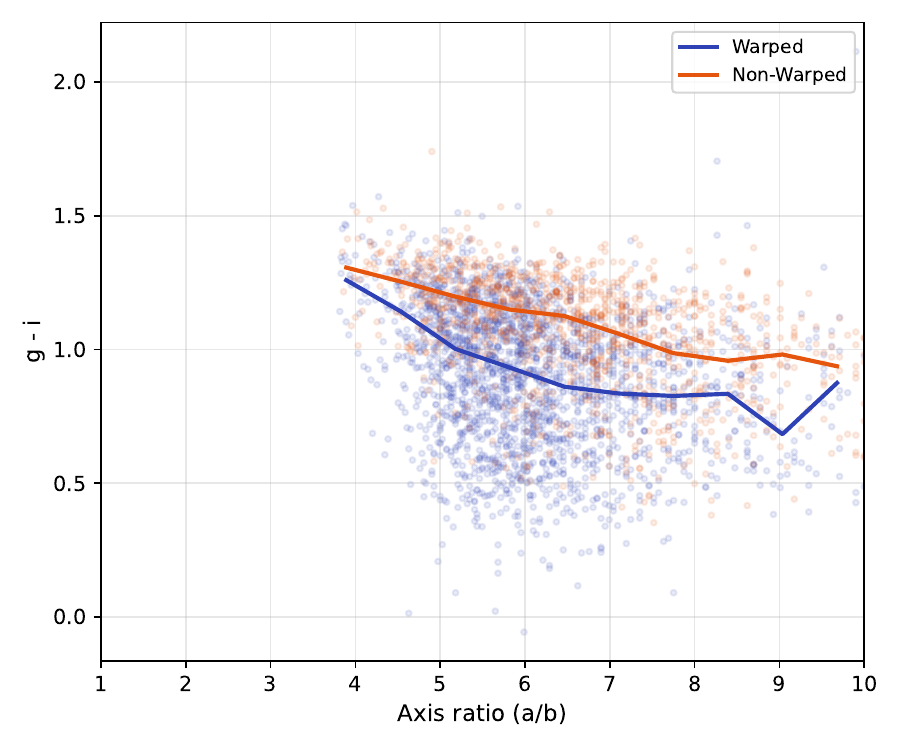}
    \caption{Left: Cumulative distribution functions of $(g-i)$ colour for warped and non-warped galaxies. Right: $(g-i)$ colour as a function of axis ratio ($a/b$). The solid lines show the median trends for each population.}
    \label{fig:pan_color}
\end{figure*}

%--------------------------------------------------------------
\begin{figure*}
    \centering
    \includegraphics[width=0.45\textwidth]{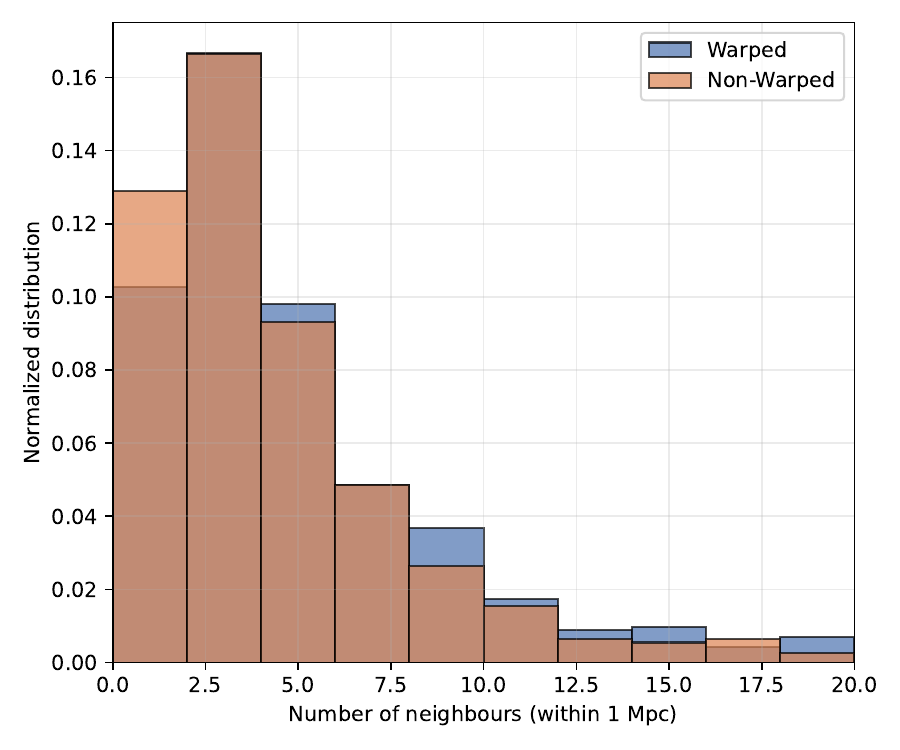}
    \includegraphics[width=0.45\textwidth]{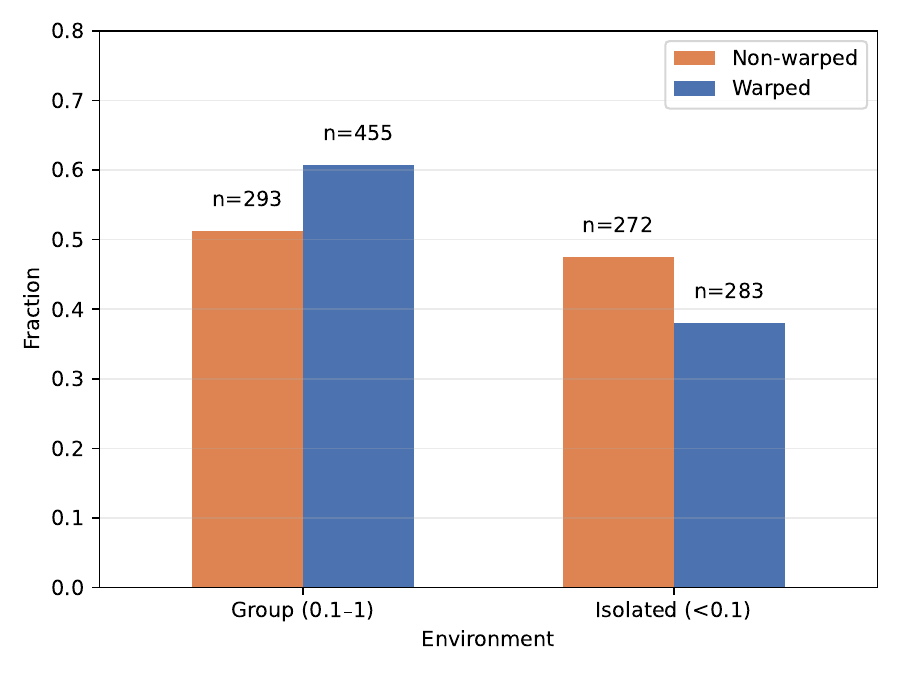}
    \caption{Left: Normalised distribution of the number of neighbours within 1 Mpc ($N_{\mathrm{1Mpc}}$) for warped and non-warped galaxies. Right: Fraction of galaxies in different environments (isolated and group) for the two populations.}
    \label{fig:pan_env}
\end{figure*}

The redshift distribution in Figure~\ref{fig:pan_z_dist} reveals an imbalance between the two populations at low redshift (0.01 $\leq$ z $<$ 0.05), while the higher-redshift range is comparatively well sampled. A similar redshift-driven bias is reported by \citet{2025arXiv250519583S}, which shows that Zoobot predicted a higher number of lopsided spiral galaxies at lower redshifts. To correct for this bias, we adopt a similar stratified subsampling approach. The low-redshift range is divided into bins of width $\Delta z = 0.01$, and within each bin the majority class is randomly subsampled to match the minority class, while all galaxies are retained at higher redshift. This procedure is repeated 100 times, and the median values are computed. All statistical analyses presented below are based on this redshift bias-corrected sample, ensuring that the inferred trends are not driven by selection effects.

\vspace{5mm}

\noindent\textit{Physical properties of warped and non-warped galaxies:}

\vspace{2mm}

\noindent We analyse several structural and physical parameters for warped and non-warped galaxies from the Pan-STARRS EGIPS catalogue. These parameters are obtained from the photometry table by combining SExtractor measurements with morphological statistics computed with StatMorph. The distributions are shown as cumulative density functions, with dashed lines indicating the median values of each class. The Kolmogorov-Smirnov (KS) test is used to assess whether two populations differ statistically. Figure~\ref{fig:pan_morph} compares the distributions of key morphological parameters for warped and non-warped galaxies. The axis ratio ($b/a$) shows a clear shift towards lower median values for warped galaxies (KS $p \sim 10^{-39}$), indicating a systematic difference in their projected shapes. The asymmetry parameter also exhibits a higher median for warped galaxies (KS $p \sim 10^{-7}$), suggesting that warped structures contribute to increased asymmetry. The concentration index shows a statistically significant difference (KS $p \sim 10^{-6}$), although the median values and overall distributions remain broadly similar. Overall, the most pronounced differences are seen in projected shape and asymmetry, while the concentration distributions are comparatively similar despite being statistically distinct.

Additionally, we cross-match the Pan-STARRS EGIPS galaxy sample with the Sloan Digital Sky Survey (SDSS) DR17 catalogue using SciServer CasJobs to obtain spectroscopic and derive physical properties that are not available in Pan-STARRS. The cross-match is performed using sky coordinates (RA, Dec), associating each Pan-STARRS galaxy with the nearest SDSS source within a 3 arcsec radius. Only spectroscopic objects classified as galaxies with reliable redshift measurements are retained. This provides a subset of galaxies with SDSS matches for further analysis. Figure~\ref{fig:pan_sdss} compares the distributions of stellar mass, star formation activity, and age for warped and non-warped galaxies. The stellar mass distribution shows a modest shift towards higher values for warped galaxies (KS $p \sim 10^{-3}$). The specific star formation rate (sSFR) is higher in warped galaxies (KS $p \sim 10^{-5}$), indicating that they exhibit enhanced star formation activity. The stellar age distribution indicates that warped galaxies are, on average, younger than non-warped galaxies (KS $p \sim 10^{-4}$), possibly due to enhanced star formation. Overall, these results indicate systematic differences between the two populations, with warped galaxies exhibiting higher star-formation activity and younger stellar ages. However, the magnitude of these shifts remains small compared to the overall distributional spread.

Similar to \citet{2022MNRAS.511.3063M}, we examine the optical colours of the galaxies using the $(g-i)$ colour derived from Pan-STARRS photometry. Figure~\ref{fig:pan_color} shows the distribution of $(g-i)$ colour for warped and non-warped galaxies. Warped galaxies exhibit lower median $(g-i)$ values (i.e. are bluer) than non-warped galaxies (KS $p \sim 10^{-15}$), indicating a statistically significant difference consistent with younger stellar populations or enhanced star formation activity. To investigate the role of projection effects, we further examine $(g-i)$ colour as a function of axis ratio ($b/a$). Both populations show a general trend of becoming bluer at lower axis ratios; however, warped galaxies remain systematically bluer than non-warped galaxies at fixed $b/a$. This suggests that the observed colour difference is not solely driven by inclination effects, but reflects an intrinsic difference between the two populations.

To investigate the role of environment in shaping galactic warps, we quantify the local galaxy density and environmental classification of our sample following the methodology of \citet{2018MNRAS.476.4488H}. We obtain environmental information from the NASA/IPAC Extragalactic Database (NED) environment service, which provides neighbour counts ($N$) and corresponding number densities ($\rho$) within fixed projected radii of 0.5, 1, 2, 5, and 10 Mpc, within a velocity window of $\pm 500$ km s$^{-1}$ around each galaxy. In particular, we focus on the number of neighbours within 1 Mpc ($N_{\mathrm{1Mpc}}$), which serves as a proxy for the local environment. The large-scale environment is characterised using the density within 5 Mpc ($\rho_{\mathrm{5Mpc}}$), based on which galaxies are classified into three categories: isolated ($\rho_{5} < 0.1$), group ($0.1 \leq \rho_{5} < 1$), and cluster ($\rho_{5} \geq 1$), consistent with the scheme adopted by \citet{2018MNRAS.476.4488H}. We compare the environmental properties of warped and non-warped galaxies using both the distribution of neighbour counts and cumulative distribution functions (CDFs). Figure~\ref{fig:pan_env} compares the environments of warped and non-warped galaxies. The neighbour counts within 1 Mpc and the fractions in isolated and group environments are similar for both populations, indicating no strong environmental dependence of warps.

%---------------------------------------------

\subsection{Prediction on Euclid Q1 images}

\begin{figure*}
    \centering
    \includegraphics[width=0.32\textwidth]{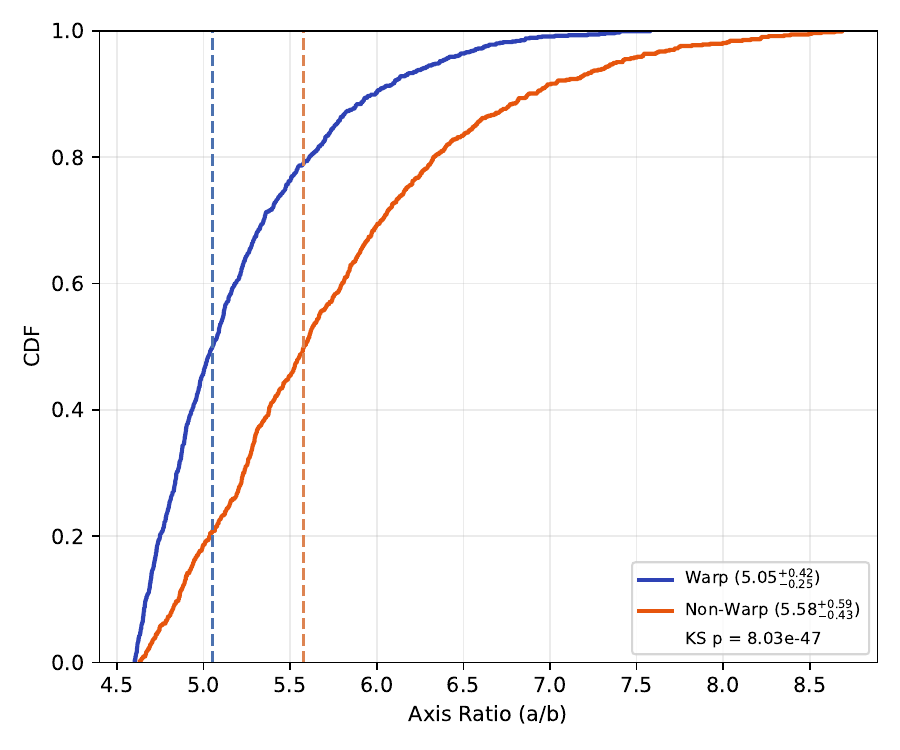}
    \includegraphics[width=0.32\textwidth]{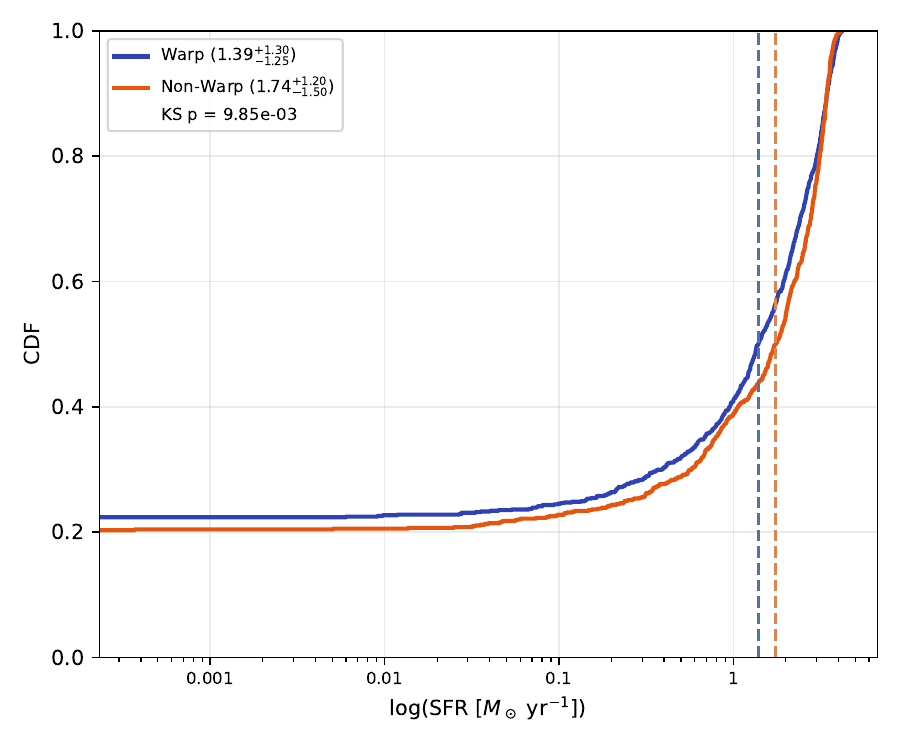}
    \includegraphics[width=0.32\textwidth]{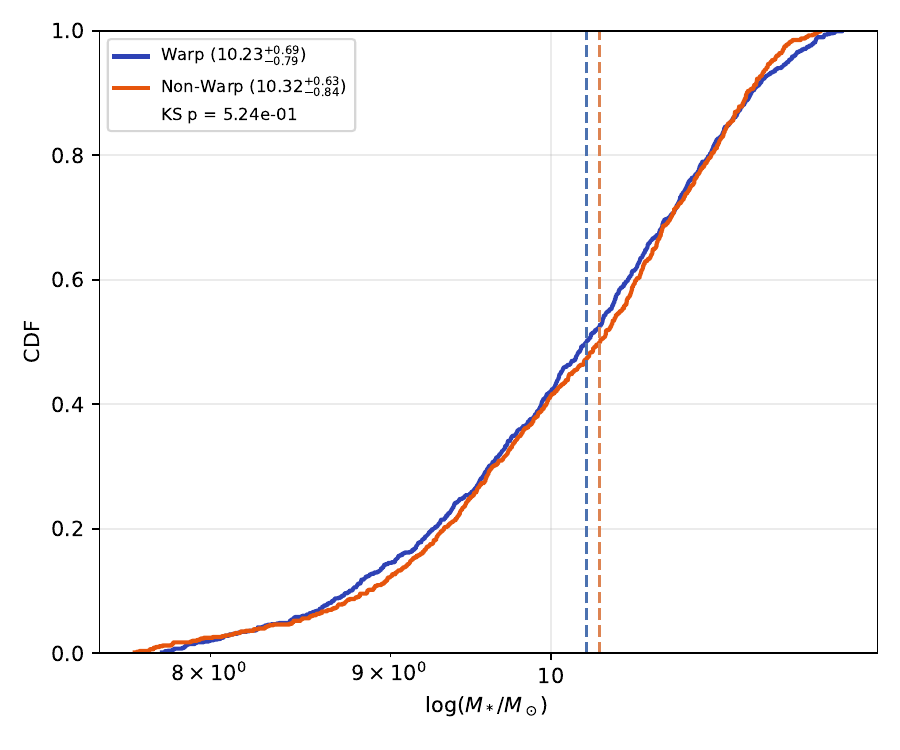}
    \caption{Cumulative distribution functions of morphological and physical properties for warped and non-warped galaxies in the Euclid sample. The dashed vertical lines indicate the median values of each population, and the KS $p$-values quantify the statistical significance of the differences. From left to right: axis ratio ($a/b$), star formation rate (SFR), and stellar mass.}
    \label{fig:euclid_props}
\end{figure*}

%------------------------------------------------

Transfer learning provides an effective framework for applying convolutional
neural networks trained on one survey to data from another, despite differences in image characteristics such as resolution, depth, and point spread function \citep{2019MNRAS.484...93D}. This reflects the ability of these models to
capture general morphological features of galaxies rather than survey-specific imaging properties. With upcoming deep surveys such as Euclid \citep{2025arXiv250315302E}, a larger number of faint
structural features, including disk warps, can be identified. Morphological
measurements on the Euclid Q1 data have been carried out by
\citet{2025arXiv250315310E}, providing a comprehensive catalogue of galaxies
across several morphological classes. However, a dedicated classification for
warped galaxies is not included. We identify warped galaxies in the sample using our
best-performing model and investigate their physical properties and
environments. The Euclid morphology catalogue, consisting of 3,80,111 galaxies,
is first filtered to select sources with a high edge-on probability
($\geq 90\%$), resulting in 14,083 galaxies. These are further divided based on
inclination into ``Gold" ($i = 90^\circ$, 2086 galaxies),
``Silver" ($85^\circ \leq i < 90^\circ$, 1140 galaxies), and
``Other" ($i < 85^\circ$, 10857 galaxies). We focus on galaxies with $i \geq 80^\circ$. This catalogue is
then cross-matched with the OU-MER source catalogue \citep{2025arXiv250315305E} to obtain the corresponding
VIS-band image cutouts from the ESA Euclid data labs. A careful visual
inspection is performed to remove unsuitable images affected by projection
effects, yielding a final sample of 3226 nearly edge-on galaxies used for
prediction and analysis. Among these, 1209 are identified as warped and 1025 as non-warped, with high prediction probabilities ($\geq 0.85$). A summary of the prediction probabilities and the corresponding number of images is given in  Table~\ref{tab:euclid_prediction_summary}.
\begin{table}
\centering
\caption{Summary of the prediction probabilities and the corresponding number of images in each class for Euclid Q1 prediction sample.}
\label{tab:euclid_prediction_summary}
\begin{tabular}{ccc}
\hline
Prediction Probability & Unwarped & Warped \\
\hline
90\% and above & 859 & 1041 \\
85-90 \% & 166 & 168 \\
85\% and below & 447 & 545 \\
\hline
\end{tabular}
\end{table} Figure~\ref{fig:euclid_highconf} presents a sample subset of the high confidence warped and non-warped systems. Euclid VIS images are deeper than Pan-STARRS imaging (by $\sim$1--1.5 mag), and therefore show more background objects within each image cutout. The redshift distribution of the Euclid sample shows no significant imbalance between warped and non-warped galaxies, and therefore no bias correction is applied, unlike in the Pan-STARRS sample. Figure~\ref{fig:euclid_props} compares the distributions of key properties for warped and non-warped galaxies in the Euclid sample. The axis ratio ($b/a$) shows a clear shift towards lower values for warped galaxies (KS $p \sim 10^{-47}$), indicating a highly significant difference consistent with the trend observed in the Pan-STARRS sample. The star formation rate (SFR) shows a statistically significant difference between the two populations (KS $p \sim 10^{-2}$), with warped galaxies exhibiting lower median values. In contrast, the stellar mass distributions are indistinguishable (KS $p \sim 0.5$), indicating no statistically significant difference between warped and non-warped galaxies. Thus, both the Pan-STARRS and Euclid samples show consistent behaviour, with the most prominent differences between warped and non-warped galaxies seen in structural properties, particularly the axis ratio. This consistency across both datasets suggests that the model captures warp-related features and yields comparable results across surveys.

%\FloatBarrier
\section{Conclusions}
\label{sec:Conclusion}
In this work, we develop a supervised deep learning framework based on a pretrained Zoobot encoder, finetuned to identify warped disc galaxies in large imaging surveys. Using edge-on galaxy FITS cutouts from the Pan-STARRS EGIPS catalogue, we construct a training sample of 1000 galaxies (500 warped and 500 non-warped), selected based on measured warp angles and visual inspection. The model achieves a mean test accuracy of $(83 \pm 2.24)\%$ with an AUC of 0.90 across 10 independent runs. The best-performing model, with a test accuracy of 87\%, was applied to a larger sample of 5,393 edge-on galaxies, identifying 2,088 warped and 1,398 non-warped systems at a prediction threshold of 85\%, thereby constructing a high-confidence catalogue. The model was further evaluated on 3226 edge-on galaxies from the Euclid Q1 survey, of which 1209 were predicted to be warped, and 1025 were non-warped, with high-confidence prediction probabilities ($\geq 0.85$). We further analyse the morphological and physical properties of the identified galaxies. The results show that warped galaxies differ primarily in structural parameters, with lower axis ratios and higher asymmetry than non-warped systems. In contrast, their physical properties, including stellar mass and environment, show no significant differences. Warped galaxies, however, are systematically bluer, with younger stellar ages and higher star-formation activity. The consistency of these trends across the Pan-STARRS and Euclid samples indicates that warp-related features are robustly recovered across surveys with different depths and resolutions, demonstrating the effectiveness of deep learning methods for analysing large imaging datasets.

\section*{Acknowledgements}

We thank Mr Biju Saha for useful discussions and suggestions.

This work uses the publicly available Pan-STARRS EGIPS catalogue, derived from the Pan-STARRS1 Surveys (PS1), which provides photometric and morphological measurements for a large sample of galaxies. The Pan-STARRS1 Surveys and the PS1 public science archive have been made possible through contributions by the Institute for Astronomy, the University of Hawai‘i, the Pan-STARRS Project Office, the Max Planck Society and its participating institutes, the Max Planck Institute for Astronomy and the Max Planck Institute for Extraterrestrial Physics, The Johns Hopkins University, Durham University, the University of Edinburgh, Queen’s University Belfast, the Harvard–Smithsonian Center for Astrophysics, the Las Cumbres Observatory Global Telescope Network Incorporated, the National Central University of Taiwan, the Space Telescope Science Institute, the National Aeronautics and Space Administration under Grant No. NNX08AR22G issued through the Planetary Science Division of the NASA Science Mission Directorate, the National Science Foundation under Grant No. AST-1238877, the University of Maryland, Eötvös Loránd University (ELTE), the Los Alamos National Laboratory, and the Gordon and Betty Moore Foundation.

We thank the Sloan Digital Sky Survey (SDSS) for making their data products publicly available, and the SciServer CasJobs platform for enabling access and queries. We also make use of the NASA/IPAC Extragalactic Database (NED), operated by the California Institute of Technology and funded by NASA.

This paper includes data from the Euclid mission, a European Space Agency (ESA) mission with contributions from ESA member states and NASA. We thank the Euclid Consortium and the Euclid data processing centres. We also acknowledge the Euclid Quick Data Release (Q1) visual morphology catalogue presented in \citet{2025arXiv250315310E}, along with the associated data products described in \citet{2025arXiv250315305E}.

\section*{Data Availability}

The data used in this study are publicly available through our GitHub repository\footnote{\url{https://github.com/saranya-suguna/Warped_DCNN}}. and codes 
will be made available with a reasonable request to the corresponding 
author. 


\bibliographystyle{mnras}
\bibliography{references}

%%%%%%%%%%%%%%%%%%%%%%%%%%%%%%%%%%%%%%%%%%%%%%%%%%
% Don't change these lines
\bsp	% typesetting comment
\label{lastpage}
\end{document}